\documentclass{aa}
\usepackage{epsf}

\begin{document}

% *************************************************************************
%                             ABBREVIATIONS
% *************************************************************************
\def\pFn{p_{\raise-0.3ex\hbox{{\scriptsize F$\!$\raise-0.03ex\hbox{\rm n}}}} }  % p_Fn
\def\pFp{p_{\raise-0.3ex\hbox{{\scriptsize F$\!$\raise-0.03ex\hbox{\rm p}}}} }  % p_Fp
\def\pFe{p_{\raise-0.3ex\hbox{{\scriptsize F$\!$\raise-0.03ex\hbox{\rm e}}}} }  % p_Fe
\def\pFmu{p_{\raise-0.3ex\hbox{{\scriptsize F$\!$\raise-0.03ex\hbox{\rm $\mu$}}}} }  % p_Fe
\def\m@th{\mathsurround=0pt }
\def\eqalign#1{\null\,\vcenter{\openup1\jot \m@th
   \ialign{\strut$\displaystyle{##}$&$\displaystyle{{}##}$\hfil
   \crcr#1\crcr}}\,}
\newcommand{\vp}{\mbox{\boldmath $p$}}         % vector p (momentum)
\newcommand{\vS}{\mbox{\boldmath $S$}}
\newcommand{\vP}{\mbox{\boldmath $P$}}

\newcommand{\om}{\mbox{$\omega$}}              % \omega
\newcommand{\Om}{\mbox{$\Omega$}}              % \Omega
\newcommand{\Th}{\mbox{$\Theta$}}              % \Theta
\newcommand{\ph}{\mbox{$\varphi$}}             % \varphi
\newcommand{\del}{\mbox{$\delta$}}             % \delta
\newcommand{\Del}{\mbox{$\Delta$}}             % \Delta
\newcommand{\lam}{\mbox{$\lambda$}}            % \lambda
\newcommand{\Lam}{\mbox{$\Lambda$}}            % \Lambda
\newcommand{\ep}{\mbox{$\varepsilon$}}         % \varepsilon
\newcommand{\ka}{\mbox{$\kappa$}}              % \kappa
\newcommand{\dd}{\mbox{d}}                     % d - differential
\newcommand{\vect}[1]{\bf #1}                % vector (жирный прямой)
\newcommand{\vtr}[1]{\mbox{\boldmath $#1$}}  % vector (жирный наклонный)
\newcommand{\vF}{\mbox{$v_{\mbox{\raisebox{-0.3ex}{\scriptsize F}}}$}}  % v_F
\newcommand{\pF}{\mbox{$p_{\mbox{\raisebox{-0.3ex}{\scriptsize F}}}$}}  % p_F
\newcommand{\kF}{\mbox{$k_{\rm F}$}}           % k_F
\newcommand{\kTF}{\mbox{$k_{\rm TF}$}}         % k_TF
\newcommand{\kB}{\mbox{$k_{\rm B}$}}           % k_B
\newcommand{\tn}{\mbox{$T_{{\rm c}n}$}}        % Tcn (crit.tem-re of neutrons)
\newcommand{\tp}{\mbox{$T_{{\rm c}p}$}}        % Tcp (crit.tem-re of protons)
\newcommand{\te}{\mbox{$T_{eff}$}}             % T effective
\newcommand{\ex}{\mbox{\rm e}}                 % exponent (roman type)
\newcommand{\rate}{\mbox{${\rm erg~cm^{-3}~s^{-1}}$}}
\newcommand{\mur}{\raisebox{0.2ex}{\mbox{\scriptsize (э)}}} %--------------%
\newcommand{\Mn}{\raisebox{0.2ex}{\mbox{\scriptsize (э{\it n\/})}}}        %
\newcommand{\Mp}{\raisebox{0.2ex}{\mbox{\scriptsize (э{\it p\/})}}}        %
\newcommand{\MN}{\raisebox{0.2ex}{\mbox{\scriptsize (э{\it N\/})}}}        %

% *************************************************************************
%                                 TITLE
% *************************************************************************
\title{ \bf Neutrino Emission from Superfluid Neutron-Star
            Cores: Various Types of Neutron Pairing}

\author{ M.~E.\ Gusakov   }
\institute{
         Ioffe Physical Technical Institute,
         Politekhnicheskaya 26, 194021 St.~Petersburg, Russia\\
     {\it e-mail: gusakov@astro.ioffe.rssi.ru }
}
\offprints{M.E.\ Gusakov}
\date{Received 12 march 2002 / Accepted 16 april 2002}
% **********************************************************************
\abstract
{
We calculate and provide analytic fits of
the factors which describe the reduction of
the neutrino emissivity of modified Urca 
and nucleon-nucleon bremsstrahlung processes
by superfluidity of neutrons and protons in neutron-star cores.
We consider $^1$S$_0$ pairing of protons and
either $^1$S$_0$
or $^3$P$_2$ pairing of neutrons. We analyze two types of
$^3$P$_2$ pairing: the familiar pairing
with zero projection
of the total angular momentum of neutron pairs
onto quantization axis, $m_J=0$; and the pairing
with $|m_J|=2$ which leads to the gap with
nodes at the neutron Fermi surface.
Combining the new data with those available
in the literature we fully describe
neutrino emission by nucleons from neutron
star cores to be used
in simulations of cooling of superfluid neutron stars.
 \bf Neutrino Emission from Superfluid Neutron-Star
            Cores: Various Types of Neutron Pairing}
\titlerunning{Neutrino emission from superfluid neutron-star cores}
\authorrunning{M.~E.~Gusakov}
\maketitle

% *************************************************************************
%                               TEXT BODY
% *************************************************************************

%%%%%%%%%%%%%%%%%%%%%%%%%%%%%%%%%%%%%%%%%%%%%%%%%%%%%%%%%%%%%%%%%%%%%%%%%%%%%
%**************** Section 1 *******************************
\section{Introduction}

It is well-known (e.g., Yakovlev et al. \cite{yls99,ykgh01})
that cooling of neutron stars depends on the
properties of matter 
in the neutron star cores. In spite of great progress
in understanding the neutron-star physics,
many properties of this matter are still known with large uncertainty.
A comparison of the theoretical
cooling models with observations of thermal emission
from isolated neutron stars gives a potentially powerful
method to explore the internal structure of neutron stars.
For a successful modeling of the cooling one needs
reliable values of neutrino emissivity
in different neutrino reactions.

In this paper we consider the matter 
of neutron star cores 
(at densities $\rho \ga 1.5 \times 10^{14}$ g cm$^{-3}$)
composed
of neutrons (n), protons (p), and
electrons (e).
It is generally agreed the neutrons and protons
can be in superfluid state
(as reviewed, e.g., by Lombardo \& Schulze, \cite{ls01}).
Superfluidity affects the neutrino emission 
and thus the cooling of neutron stars.
According to numerous
microscopic calculations, the proton pairing
occurs in the singlet ($^1$S$_0$) state of proton pairs.
Following Yakovlev et al.\ (\cite{yls99}) we will call
this pairing as pairing A.
The neutron pairing occurs
either in the $^1$S$_0$ state 
or in the triplet state
($^3$P$_2$). Neutron pairing A takes place
in the matter of subnuclear density
($\rho \la \rho_0$, 
where $\rho_0 = 2.8 \times 10^{14}$ g cm$^{-3}$ 
is the saturated nuclear matter density), 
while the $^3$P$_2$ pairing
is efficient at higher $\rho$.
We consider the $^3$P$_2$ pairing of two types denoted
as B and C. Pairing B occurs
in a state of a neutron pair with zero
projection of the
total angular momentum on the
quantization axis, $m_J=0$. This pairing has been studied in the
majority of papers devoted to $^3$P$_2$
pairing of neutrons. Pairing C occurs
in a state with $|m_J|=2$. It has been the subject
of some studies 
(as reviewed, e.g., by Yakovlev et al.\ \cite{yls99}).
The actual type of neutron pairing (A, B, or C)
corresponds to the state with minimum free energy.
Pairing C seems to be less realistic than B
but cannot be completely ruled out by
contemporary microscopic theories.
For example, Muzikar et al.\ (\cite{mss80}) showed that
it realizes in matter with strong magnetic field ($B \ga 10^{16}$ G). 
Amundsen \& {\O}stgaard (\cite{ao85}) found that
the energetically preferable state of the pair can be 
a superposition of states with different $m_J$.  
The specific feature of pairing C is that it leads to superfluid
gaps with nodes at the neutron Fermi surface producing
qualitatively different effect on
neutrino processes than pairing B (or A). 

Note that we do not consider another case:
$^3$P$_2$ neutron pairing with $|m_J| = 1$.
In this case, just as in cases A and B,
the superfluid gap does not
have any nodes at the Fermi surface (e.g.,
Amundsen \& {\O}stgaard \cite{ao85}).
Therefore, we expect that the results will be
similar to those for pairing B or A.
On the other hand, the consideration of the $|m_J|=1$ pairing
is technically much more complicated since the superfluid
gap depends not only on the polar angle $\vartheta$ of
neutron momentum at the Fermi sphere (see below)
but also on the azimuthal angle $\varphi$. 

Let us remind five main neutrino generation mechanisms 
in the neutron-star cores. 

(1) Direct Urca process is the most powerful neutrino process. 
It consists of two successive
reactions
\begin{equation}
 {\rm n \to p+e+\bar{\nu}_{\rm e}, \quad p+e \to n + \nu_{\rm e}},
\label{urca}
\end{equation}
where $\nu_{\rm e}$ and $\bar{\nu}_{\rm e}$
are electron neutrino and antineutrino.
It is allowed (Lattimer et al.\ \cite{lpph91})
only in matter of sufficiently high density
(typically, a few times of $\rho_0$) for model equations
of state with high symmetry energy
(rather high fraction of protons).
The reduction of direct Urca process by
proton superfluidity A and neutron
superfluidity (A, B, or C) was analyzed
by Levenfish \& Yakovlev (\cite{ly94}).

(2) Modified Urca process consists of two branches.
Two successive reactions (direct and inverse)
\begin{equation}
  {\rm n+n \to p+n+e+\bar{\nu}_e, \quad p+n+e \to n+n + \nu_e}
\label{Modn}
\end{equation}
form the {\it neutron branch}. Two similar reactions
\begin{equation}
  {\rm n+p \to p+p+e+\bar{\nu}_e, \quad p+p+e \to n+p + \nu_e}
\label{Modp}
\end{equation}
form the {\it proton branch} of the process.
The process is the most powerful
neutrino mechanism in non-superfluid 
neutron-star cores where the direct Urca process
is forbidden. It (or at least
its neutron branch) is open in the entire stellar
core.

The neutrino emissivity of 
this process in non-superfluid matter was considered
by a number of authors (references can be found
in Yakovlev et al.\ \cite{yls99}), particularly, by
Bahcall \& Wolf (\cite{bw65}), Friman \& Maxwell (\cite{fm79}),
and Yakovlev \& Levenfish (\cite{yl95}). 
The latter authors studied the reduction
of the process either by 
proton superfluidity A, or
by neutron superfluidity (A or B).
Levenfish \& Yakovlev (\cite{ly96})
suggested a simple approximate method
to account for the combined effect
of the neutron and proton superfluidities.
It is based on the similarity relations of the
factors which describe the superfluid reduction
of the direct and modified Urca processes.
These results were used in simulations
of the neutron star cooling
(as reviewed by Yakovlev et al.\ \cite{yls99,ykgh01}).
We present a more accurate calculation
of the reduction of the modified
Urca process by combined action of proton
superfluidity A and neutron superfluidity (A, B, or C).

(3) The neutrino-pair bremsstrahlung
at nucleon-nucleon scattering can be of three types:
\begin{eqnarray}
{\rm n + n} &\to& {\rm n + n + \nu + \bar{\nu}},
\label{brems_nn} \\
{\rm n + p} &\to& {\rm n + p + \nu + \bar{\nu}},
\label{brems_np} \\
{\rm p + p} &\to& {\rm p + p + \nu + \bar{\nu}},
\label{brems_pp} 
\end{eqnarray}
Here, $\nu$ and $\bar{\nu}$ stand for neutrinos of
any flavor. In a normal (non-superfluid) matter
the bremsstrahlung processes are weaker
than the modified Urca process 
(e.g., Yakovlev et al. \cite{yls99}).
However they may be more important in superfluid matter.
If proton superfluidity is of type A, and neutron superfluidity
is of type A or B then superfluid reduction of
the processes
can be described by the formulae presented by
Yakovlev et al.\ (\cite{yls99}).
In this paper we develop analogous description
for neutron superfluidity C.

(4) Neutrino emission due to Cooper pairing
of nucleons (N= n or p) actually consists of neutrino-pair
(any flavor) emission 
\begin{equation}
     {\rm N \to N + \nu + \bar{\nu}}
\label{coop}
\end{equation}
by a nucleon 
whose dispersion relation
contains an energy gap.
The process was proposed by Flowers et al.\ (\cite{frs76})
for neutron superfluidity of type A.
The extension to the neutron superfluidity
B and C was done by Yakovlev et al.\ (\cite{ykl99}).
The case of proton superfluidity A is described, e.g.,
by Yakovlev et al.\ (\cite{yls99,ykgh01}).
In the absence of superfluidity, the reaction
is forbidden by energy-momentum conservation.

(5) Neutrino-pair bremsstrahlung at
electron-electron scattering (Kaminker \& Haensel \cite{kh99}),
\begin{equation}
  {\rm e + e \to e + e + \nu + \bar{\nu}},
\label{brems_ee}
\end{equation}
is much weaker than
other processes in non-superfluid matter.
However, it is almost independent of superfluidity
and may be the leading mechanism in
superfluid matter.

The present paper is organized as follows.
In Sect.\ 2 we present general equations
for modified Urca process and analyze
the reduction
factors.
In Sect.\ 3 we consider the reduction factors
of nucleon-nucleon bremsstrahlung processes.
In Sect.\ 4 we study the efficiency of various
neutrino processes in the cores of neutron
stars for different superfluidity types.
Analytic fits of the reduction factors of
the modified Urca process are given in Appendix.

%%%%%%%%%%%%%%%%%%%%%%%%%%%%%%%%%%%%%%%%%%%%%%%%%%%%%%%%%%%%
\section{Modified Urca process}
%%%%%%%%%%%%%%%%%%%%%%%%%%%%%%%%%%%%%%%%%%%%%%%%%%%%%%%%%%%%
%
\subsection{General equations}
As discussed,
e.g., by Bahcall \& Wolf (\cite{bw65}) and Friman \& Maxwell (\cite{fm79}),
the general expression for the neutrino emissivity
of modified Urca process
can be written as
($\hbar = c = \kB = 1$):
\begin{eqnarray}
    Q & = & 2  \int
              \left[ \prod_{j=1}^4 { \dd^3 p_j \over (2 \pi)^3} \right]
              { \dd^3 p_{\rm e} \over 2 \ep_{\rm e} (2 \pi)^3} \;
              { \dd^3 p_\nu \over 2 \ep_\nu (2 \pi)^3} \;  \ep_\nu \,
\nonumber \\
      & & \times \;
       (2 \pi)^4 \, \delta(E_f-E_i) \, \delta({\vec{P}}_f - \vec{P}_i ) \,
               { {\cal L}\over 2} \, \sum_{\rm spins} | M |^2 ,
  \label{eq:Q_mur}
\end{eqnarray}
where 
${\vec{p}}_j$ is a nucleon momentum
($j=1$,~2,~3,~4); ${\vec{p}}_{\rm e}$ and
$\ep_{\rm e}$ are, respectively, the momentum and energy of
an electron; and
${\vec{p}}_\nu$ and $\ep_\nu$ are the momentum and energy
of a neutrino. The delta function
$ \del(E_f-E_i) $ describes energy conservation,
while $\del( {\vec{P}}_f - {\vec{P}}_i )$
describes momentum conservation. The indices $i$ and $f$
refer to the initial and final particle states.
Furthermore, ${\cal L}$ means the product of
Fermi-Dirac functions or corresponding blocking
factors of the nucleons and the electron;
$ | M |^2 $ is the squared
matrix element. Summation is carried over
all particle spins. The factor 2 in the denominator
before the summation sign is introduced to avoid
double counting of the same reactions involving
identical particles. The overall factor 2 
doubles the emissivity of elementary (direct or inverse)
reaction of the process assuming beta-equilibrium.
Since neutrons, protons and electrons in neutron-star
cores are strongly degenerate, one can 
use the phase-space decomposition (e.g., Friman \& Maxwell \cite{fm79})
which yields:
\begin{eqnarray}
&&    Q  =  {1 \over 4 \, (2 \pi)^{14} }\; T^8
            AI \; \prod_{j=1}^5 \pF_{\!j} m_j^\ast
            \; \sum_{\mbox{\scriptsize spins}} | M |^2 ,
\label{eq:DecompMur} \\
&&   A = 4 \pi \;  \left[ \prod_{j=1}^5 \int \dd \Om_j \; \right]
            \del \left( \sum_{j=1}^5 {\vec{p}}_j  \right),
\label{eq:A} \\
&&   I = \int_0^\infty \! \! \dd x_\nu \, x_\nu^3
       \left[ \prod_{j=1}^5 \int_{-\infty}^{+\infty}
       \dd x_j 
       %\, 
       f_j \right]
       \del \left( \sum_{j=1}^5 x_j-x_\nu \right) \! .
\label{eq:I}
%\\
%&&   S = \prod_{j=1}^5 \pF_{\!j} m_j^\ast.
%\label{eq:S}
\end{eqnarray}
The factor $A$ contains integrations over orientation
of particle momenta ($j=5$ refers to an electron);
$ \dd \Om_j$ is a solid angle element in the direction of
${\vec{p}}_j$. All lengths of the momenta ${\vec{p}}_j$
with $j \leq 5$ can be set equal to the appropriate
Fermi momenta $p_{\rm F}$. 
A typical neutrino momentum $p_\nu$
is determined by the temperature $T$, $p_\nu \sim T \ll p_{\rm F}$.
Thus, we neglect $p_\nu$ in the momentum-conserving
delta function and integrate over orientations
of $\vec{p}_\nu$ in $A$ (which gives a factor of  $4 \pi$).
The factor $I$ given by Eq.\ (\ref{eq:I})
contains the integrals over the dimensionless
neutrino energy $x_\nu = p_{\nu}/T =\ep_\nu /T$
and the dimensionless energies of other particles
$x_j = \vF_{\!j} (p-\pF_{\!j})/T$;
$\; f_j= [ \exp(x_j)+1]^{-1}$. Finally, Eq.\ (\ref{eq:DecompMur})
contains the products of the densities of state
of the particle species $1 \le j \le 5$,
$m_j^\ast$ being the effective mass at the Fermi surface.
For non-relativistic nucleons (which we consider here), $m_j^\ast$
is mainly determined by in-medium effects.
For the electrons ($j$=5), $m^\ast=\mu$,
where $\mu$ is the electron chemical potential.
In the absence of superfluidity Eq.\ (\ref{eq:I})
gives $I=I_{0}= 11513 \, \pi^8\,/ 120960$.
For the neutron branch (\ref{Modn})
of the process, Eq.\ (\ref{eq:A}) yields
(e.g., Shapiro \& Teukolsky \cite{st83}):
\begin{equation}
     A^{\rm n}_{0} = {2 \pi \, (4 \pi)^4 \over \pFn^3}.
\label{An}
\end{equation}
This result is valid as long as $\pFn > \pFp + \pFe$.
Otherwise $A^{\rm n}_{0}$ is given by Eq.\ (13) in Yakovlev
\& Levenfish (\cite{yl95}) but in that case the modified Urca
process becomes insignificant because the direct
Urca process dominates.

For the proton branch (\ref{Modp}) at $\pFn \geq 3 \pFp - \pFe$
one has
\begin{equation}
     A^{\rm p}_{0} = {2 (2 \pi)^5 \over \pFn
      \pFp^3 \pFe}
       (\pFe+3 \pFp - \pFn)^2 \,  \Theta,
\label{Ap}
\end{equation}
while at  $\pFn < 3 \pFp - \pFe$ $\;$ (and $3\pFp \geq \pFe$)
\begin{equation}
     A^{\rm p}_{0} = \frac {2^3 (2 \pi)^5}{\pFp^2}
       \left( \frac {3}{\pFn} - \frac {1}{\pFp} \right) \, \Theta,
\label{Ap1}
\end{equation}
where $\Theta=1$ if the proton branch is allowed
by momentum conservation
($\pFn \leq 3 \pFp + \pFe$), and $\Theta=0$
otherwise.
Notice that Eq.\ (\ref{Ap1}) can be useful if dense
matter contains other particles but n, p, and e (e.g., muons).

The difference of Eqs.\ (\ref{An}) and (\ref{Ap}) or (\ref{Ap1})
is the consequence of the fact that $\pFn$ is significantly larger
than $\pFp$ in neutron star matter.

Combining these results one can obtain the
neutrino emissivities $Q_0^{\rm n}$ and $Q_0^{\rm p}$
in nonsuperfluid matter. The emissivity $Q_0^{\rm n}$
was calculated by Friman \& Maxwell (\cite{fm79}), using 
the one-pion-exchange
approximation for calculating the matrix element, $|M|^2$,
and $Q_0^{\rm p}$ was calculated
by Yakovlev \& Levenfish (\cite{yl95}) using the same technique.

Now consider the modified Urca process in
the presence of superfluidity of neutrons and protons.
A onset of superfluidity leads to the appearance
of an energy gap $\delta$ in the momentum dependence of
the particle energy $\varepsilon(\vp)$.
Near the Fermi surface ($| p-p_{F} | \ll p_{F}$),
this dependence can be written as
(e.g., Lifshitz \& Pitaevskii \cite{lp80})
\begin{eqnarray}
  \varepsilon & = & \mu - \sqrt{\delta ^2 + \eta ^2}
  \quad p< \pF, \quad \quad
\nonumber\\
  \varepsilon & = & \mu + \sqrt{\delta ^2 + \eta ^2} \quad p \geq \pF.
\label{epspf}
\end{eqnarray}
Here, $\eta=\vF(p-\pF)$, $\vF$ and $\pF$
are the nucleon Fermi velocity and Fermi momentum,
and $\mu$ is their chemical potential. For the conditions
of our interest, $ \delta \ll \mu $. Furthermore,
$\delta ^2= \Delta^2 (T) F(\vartheta)$, where
$\Delta(T)$ is the gap amplitude, which determines
the temperature dependence of the gap width, while
$F(\vartheta)$ is the factor which depends on the angle
$\vartheta$ between the quantization axis and the particle
momentum. The functions $\Delta(T)$ and
$F(\vartheta)$ depend on superfluidity type
(e.g. Yakovlev et al. \cite{yls99}).
For cases A, B and C:
\begin{eqnarray}
& F_{\rm A}(\vartheta) = 1,~~~~~~\quad \quad & T_{c\rm A} = 0.5669 \; \Delta(0);
 \label{FthetaA} \\
& F_{\rm B}(\vartheta) = 1+ 3 \cos ^2 \vartheta, &
 T_{c\rm B} = 0.8416 \; \Delta(0);
 \label{FthetaB} \\
& F_{\rm C}(\vartheta) = \sin ^2 \vartheta, ~\quad \quad & 
 T_{c\rm C} = 0.4926 \; \Delta(0).
\label{FthetaC}
\end{eqnarray}
The gap amplitude $\Delta$($T$)
is assumed to be governed by the standard equations of the BCS theory
(e.g., Tamagaki \cite{tamagaki70}); $\Delta$(0)
is related to the critical temperature $T_c$ as indicated above.

For further analysis it is convenient
to introduce the dimensionless variables:
\begin{eqnarray}
   z& = &\frac{ \varepsilon - \mu}{T}
    = {\rm sign} (x) \sqrt{x^2+y^2},
\nonumber \\
   x& = &\frac{\eta}{T}, \quad
   y=\frac{\delta}{ T}, \quad
   v=\frac{\Delta (T)}{T}.
 \label{bez}
\end{eqnarray}
While $T$ decreases from  $T_c$ to 0, the parameter
$v$ varies from 0 to $\infty$ as described, e.g., by 
Eq.\ (11) in Yakovlev et al.\ (\cite{yls99}).

We assume that
the neutrino emissivity in superfluid matter
can be calculated from Eqs.\ (\ref{eq:DecompMur})---(\ref {eq:I})
by replacing $x_j \to z_j$ for all
particle species which are in superfluid state.
This assumption is widely used in the literature;
its validity is discussed by Yakovlev et al.\ (\cite{ykgh01}).
In this approximation, the neutrino emissivity of the modified
Urca process can be written as
\begin{equation}
 Q^{\rm n} = Q^{\rm n}_{0} R^{\rm n}, \quad Q^{\rm p} = Q^{\rm p}_{0} R^{\rm p},
 \label{Qn}
\end{equation}
where $Q^{\rm n}_0$ and $Q^{\rm p}_0$ are the emissivities in
a non-superfluid matter, while
$R^{\rm n}$ and $R^{\rm p}$ are the factors which describe
the reduction of the emissivity by nucleon superfluidity
($R^{\rm N}< 1$). Generally, these factors
can be written as
\begin{eqnarray}
 R^{\rm N} &=& \frac{J_{\rm N}}{I^{\rm N}_0 A^{\rm N}_0},
\nonumber \\
 J_{\rm N} &=& 4 \pi \int \prod_{j=1}^5 \dd \Omega_j
          \int_0^{\infty} \dd x_\nu \, x_\nu^3
          \left[ \prod_{j=1}^5 \int_{-\infty}^{+\infty}
          \dd x_j \, f(z_j) \right]
\nonumber \\
         &\times& \del \left( x_\nu - \sum_{j=1}^5 z_j \right)
         \del \left( \; \sum_{j=1}^5 \vec{p}_j \; \right).
 \label{JN}
\end{eqnarray}
The notations are the same as in Eqs.\
(\ref{eq:DecompMur})---(\ref{eq:I}).

We have composed a code
which calculates the reduction factor (\ref{JN})
for proton superfluidity A and  
neutron superfluidity A, B, or C.
The code has been tested by comparing
with the analytical asymptotes at large
$v_1$ and $v_2$ and with the results 
of Yakovlev \& Levenfish (\cite{yl95})
who considered superfluidity of either 
protons or neutrons.
The results have also been compared with those
calculated from Eq.\ (\ref{JN}) under simplified assumption 
$\pFp$, $\pFe \ll \pFn$
discussed below (see Eq.\ \ref{RN_AB}). 

Notice that the results of this section can also be used
to describe modified Urca process with muons
instead of electrons (see Yakovlev et al.\ \cite{ykgh01},
for details).

%%%%%%%%%%%%%%%%%%%%%%%%%%%%%%%%%%%%%%%%%%%%%%%%%%%%%%%%%%%%%%%%%%%
\subsection{Reduction by
         superfluidity of neutrons and protons of type {\rm A} }
%%%%%%%%%%%%%%%%%%%%%%%%%%%%%%%%%%%%%%%%%%%%%%%%%%%%%%%%%%%%%%%%%%%

In this case Eq.\ (\ref{JN}) can be simplified.
For pairing A, the dimensionless
energy gap $y_{\rm A}$ is angle-independent.
This allows one to decompose the integrals over
the angles and over the dimensionless energies
$x_j$. For the neutron branch of the modified
Urca process we get
\begin{eqnarray}
     R^{\rm n}_{{\rm AA}} &=& \frac {120960}{11513 \pi^8}
              \left[ \prod_{j=1}^4 \int_{-\infty}^{+\infty}
              \dd x_j \, f(z_j) \right]
\nonumber \\
     & & \times
      \; G(z_1+z_2+z_3+z_4),
\label{RAn}          \\
G(a) &\equiv& \int_{-a}^{+\infty}
       \! \! \! dx \;\, \frac{(x+a)^3}{1+e^x},
\label{Ga}
\end{eqnarray}
where $j=1,2$, and $3$ enumerates the reacting neutrons,
while $j=4$ refers to a proton.

Let {\it here and hereafter} $v_1 \equiv v_{\rm n}$ 
refer to neutrons, and $v_2 \equiv v_{\rm p}$ refer to protons.
The reduction of
the proton branch is evidently given by
\begin{equation}
  R_{{\rm AA}}^{\rm p}(v_1,\, v_2) = R_{{\rm AA}}^{\rm n}(v_2,\,v_1)
\label{RAp}
\end{equation}

It is useful to find the asymptotes 
of $R_{{\rm AA}}^{\rm n}$ from Eq.\ (\ref{RAn})
for strong superfluidity, i.e., for large values of
$v_1$ and $v_2$. In this case we can introduce
three regions of the parameters (Fig.\ 1),
where the asymptote of
$R_{{\rm AA}}^{\rm n}$ has different forms. 
Region ${\rm I}$ corresponds to ${ v_{1}}>{ v_{2}}$;
region ${\rm II}$ corresponds to ${ v_{2}} \geq { v_{1}} \geq  { v_{2}}/3$;
and region ${\rm III}$ corresponds to ${ v_{1}} < { v_{2}}/3$.
The asymptotes presented below are valid at
$({ v_{1}}-{ v_{2}})^2 \gg  v_1$
and
$({ v_{2}}-3{ v_{1}})^2 \gg  v_2$, i.e.,
not too close to
the boundaries between regions I and II
and between regions
II and III.

For example, we outline the derivation of the
asymptote of $R_{{\rm AA}}^{\rm n}$ from Eq.\ (\ref{RAn})
in region I; the derivation in other regions is similar.
Clearly, the integral
(\ref{RAn})
can be subdivided into several parts in such a way that
any single part contains integrations
from $ - \infty$ to 0 and/or from 0 to +$\infty$.
Now let us introduce the convenient notations for
these parts. Let
$R(2,-1)$ mean a five-dimensional integral
containing the integration
from 0 to +$\infty$ over two neutron variables,
and over $- \infty$ to 0 over a proton variable
(in this case, the integration over the third
neutron variable is assumed to extend from
$- \infty$ to 0). Splitting the initial integral
(\ref{RAn}) into the elementary integrals,we see that the same integral
$R(2,-1)$ enters the sum three times.
Thus, it is sufficient to calculate the
integral once and multiply by 3.

In this way we obtain eight integrals of different types:
$R(3,+1)$, $R(3,-1)$, $R(2,+1)$, $R(2,-1)$, $R(1,+1)$, $R(1,-1)$, $R(0,+1)$,
and $R(0,-1)$. In the limit of strong superfluidity ($v_{1,2} \gg 1$),
each of them is exponentially small.
The exponentials are:
\begin{tabular}{ll}
                                     &                               \\
 $ \! \! R(3,+1) \propto \exp(-3 v_1- v_2),$ & $ \! \! \! \! R(1,+1)\propto \exp(-2 v_1), $ \\
 $ \! \! R(3,-1) \propto \exp(-3 v_1),     $ & $ \! \! \! \! R(1,-1)\propto \exp(-2 v_1- v_2),$\\
 $ \! \! R(2,+1) \propto \exp(-2 v_1- v_2),$ & $ \! \! \! \! R(0,+1)\propto \exp(-3 v_1), $ \\
 $ \! \! R(2,-1) \propto \exp(-2 v_1),     $ & $ \! \! \! \! R(0,-1)\propto \exp(-3 v_1- v_2).$\\
                                     &
%
%\label{exp}
\end{tabular}
%
%%%%%%%%%%%%%%%%%%%%%%%%%%%%%%%%%%%%%%%%%%%%%%%%%%%%%%%%%%%%%%
\begin{figure}[t]
\centering
\epsfysize=80mm
\epsffile[78 212 553 678]{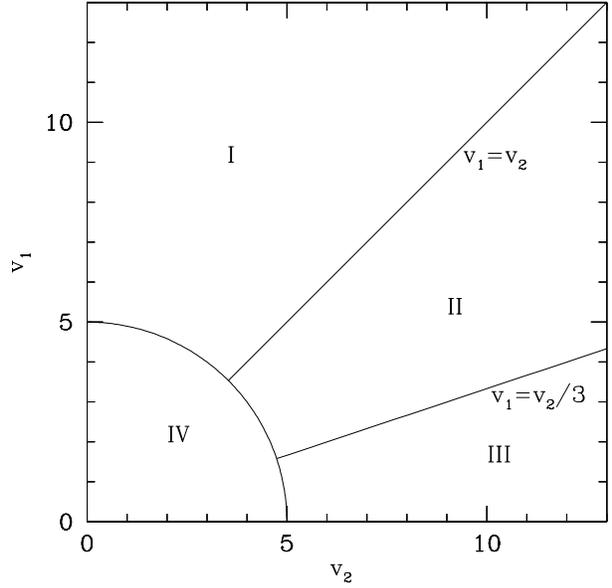}
\caption{
Four regions of $v_1 = v_{\rm n}$ and $v_2 = v_{\rm p}$
where the reduction factors $R^{\rm N}$ of the neutron
and proton branches of modified Urca process  
can be fitted by different expressions.
In regions {\rm I}, {\rm II}, and {\rm III} at $v_1 \gg 1$
and $v_2 \gg 1$ the factors $R^{\rm n}$ of neutron
branch have different asymptotes.
}
\label{area}
\end{figure}
%%%%%%%%%%%%%%%%%%%%%%%%%%%%%%%%%%%%%%%%%%%%%%%%%%%%%%%%%%%%%%%%%%%%%%

It is seen that the main contribution into
the asymptote comes from $R(2,-1)$ and $R(1,+1)$.
These terms have the same exponential but
$R(2,-1)$ has a larger pre-exponent.
Therefore, it is
$R(2,-1)$ which gives the main contribution
in region I:
\begin{eqnarray}
  R(2,-1) &=& \frac {120960}{11513 \pi^8} \int \!\! \int_0^{+\infty}
  \dd x_1 \, f(z_1)  \; \dd x_2 \, f(z_2)
\nonumber \\
   & & \, \times \,   \int \! \! \int_{-\infty}^0
  \dd x_3 \, f(z_3)  \; \dd x_4 \, f(z_4) \, G(a),
\label{Rr}
\end{eqnarray}
where $a = z_1+z_2+z_3+z_4$.
One has $G(a) \to a^4 /4$ as $a \rightarrow \infty$
and $G(a) \to 6 \exp (a)$ as $a \rightarrow -\infty$.
We are especially interested in large and positive $a$
for which $G(a) \approx a^4 \,\theta (a)/4$, where
$\theta (x)$ is the step function. In this approximation, Eq.(\ref{Rr})
can be rewritten as:
\begin{eqnarray}
  R(2,-1) &=& \frac {120960}{11513 \pi^8}  \int \int_0^{+\infty}
  \dd x_1 \, f(z_1) \; \dd x_2 \, f(z_2)
\nonumber \\
        & & \, \times \; \int \int_{-\infty}^0
  \dd x_3 \, f(z_3) \; \dd x_4 \, f(z_4)
  \frac {a^4}{4} \theta (a).
\label{Rrr}
\end{eqnarray}

Two integrals in Eq.\ (\ref{Rrr}) over the neutron variables
$x_1$ and $x_2$ are rapidly converging because of the
presence of exponentially decreasing functions.
Accordingly, the region of space, where these two
variables produce the main contribution, is rather small.
Thus, it is sufficient to set $z_1=v_1$ and $z_2= v_1$
in the $\theta$ function. Then the integrations over
$x_1$ and $x_2$ are separated and done analytically.
Neglecting exponentially small terms
while integrating over $x_3$ and $x_4$,
we obtain the asymptote of $R_{{\rm AA}}^{\rm n}$
in region I:
\begin{eqnarray}
    R_{{\rm AA}}^{\rm n} &=& 3 \; \frac{120960}{11513 \pi^8} \;
    \frac{\pi}{2} \; v_1^6 \; v_2 \; e^{-2v_1} \;
    I_1 \left(\frac{v_2}{v_1} \right) ,
\label{RAn1}  \\
    I_1(\alpha) &=& {1 \over 4} \int \int_0^{+\infty}
    \dd x_3 \, \dd x_4 \, (2-r_3- \alpha r_4)^4 \,
\nonumber \\
    & & \, \times \;
    \theta(2-r_3- \alpha r_4), \quad 0 < \alpha \leq 1,
\label{I1}
\end{eqnarray}
where $r_j= \sqrt{x_j^2+1}$.

In the same manner in region II we obtain:
\begin{eqnarray}
    R_{{\rm AA}}^{\rm n} &=& 3 \; \frac{120960}{11513 \pi^8} \;
    \frac{\pi}{2} \; v_1^{13/2} v_2^{1/2} \;
    e^{-v_1-v_2}  \; I_2 \left(\frac {v_2}{v_1} \right) \;
\nonumber \\
    &&
    + \; \frac{120960}{11513 \pi^8} \;
    \left(\frac{\pi}{2} \right)^{3/2} \; v_1^{3/2} \;
    e^{-3v_1} K ,
\label{RAn2}  \\
    I_2(\alpha) &=& {1 \over 4} \int \int_0^{+\infty}
    \dd x_1 \, \dd x_2 \, (1+ \alpha - r_1-r_2)^4 \,
\nonumber \\
     & & \, \times \;
    \theta(1+ \alpha -r_1-r_2), \quad 1 \leq \alpha \leq 3,
\label{I2}  \\
    K &=& \frac{\sqrt{9v_1^2-v_2^2}}{120} \;
          (486v_1^4+747v_1^2 \; v_2^2+16v_2^4)
\nonumber \\
      && - \frac{3}{8} \; v_1 \; v_2^2 \; (3v_2^2+36v_1^2) \;
         \ln{ \frac { 3v_1+\sqrt{9v_1^2-v_2^2} }{v_2} } .
\label{K}
\end{eqnarray}

In region III:
\begin{eqnarray}
     R_{{\rm AA}}^{\rm n} &=& \frac{120960}{11513 \pi^8} \;
     \left(\frac{\pi}{2} \right)^{1/2}
             \; v_1^3 \; v_2^{9/2} \; e^{-v_2} \;
             I_3 \left(\frac{v_2}{v_1} \right) ,
\label{RAn3} \\
     I_3(\alpha) &=& {1 \over 4} \int \!\!  \int \!\! \int_0^{+\infty}
           \dd x_1 \, \dd x_2 \, \dd x_3 \,
           \left[ 1 -\frac{1}{\alpha}(r_1+r_2+r_3) \right]^4 \,
\nonumber \\
          & & \, \times \;
          \theta (\alpha -r_1-r_2-r_3),
\label{I3}
\end{eqnarray}
where $3 \leq \alpha < \infty$.

The asymptotes themselves contain complicated integrals.
Thus we have calculated the asymptotes of the integrals
(\ref{I1}) and (\ref{I3}). In region I we have:
\begin{eqnarray}
    I_1 &=& \frac{0.027570965}{\alpha} \qquad \qquad
    \alpha \to 0 ,
\nonumber \\
    I_1 &=& 0.0785398 \; (1- \alpha)^5 \quad  \; \alpha \to 1 .
\label{AsI1}
\end{eqnarray}
The asymptote of $I_2(\alpha)$ at $\alpha \to 1$
can be determined by matching the reduction factors
at the boundary of regions I and II.

In region III we have
\begin{eqnarray}
     I_3 &=& 0.214307 \; \left(1-\frac{3}{\alpha}\right)^{11/2} \quad \; \,\;
     \alpha \to 3,
\nonumber \\
     I_3 &=& \frac{1}{840} \; \alpha^3 \qquad \qquad \qquad \quad \,  \;
     \alpha \to \infty .
\label{AsI3}
\end{eqnarray}

We have numerically calculated the functions $I_1(\alpha)$,
$I_2(\alpha)$, and $I_3(\alpha)$
and proposed analytic fits which reproduce
the results and the asymptotes
(\ref{I1}), (\ref{I2}), and (\ref{I3})
with the
maximum error less than 1$\%$.
For region I, the fit is:
\begin{eqnarray}
      I_1 &\approx& \frac{0.02757096 \; (1- \alpha ^{p_1})^5}
      { \alpha \; (1+30 \alpha ^2+60 \alpha ^4+107.186 \alpha ^6)^{p_2}},
\label{I1Ap}
\end{eqnarray}
where
$p_1 = 1.473$ and  $p_2 = 0.1684$.

For region II:
\begin{eqnarray}
     I_2 &\approx& 0.05 \; ( \alpha^{p_1}-1)^5 \;
     (\alpha^{p_2}-p_3)^2,
\label{I2Ap}
\end{eqnarray}
with $p_1 = 1.210$, $p_2 = 0.222$, and $p_3=0.215$.

For region III:
\begin{equation}
    I_3 \approx \frac{\alpha^3}{840} \;
    \left(1-\frac{9}{\alpha^2} \right)^{5.5}
    \left(1+\frac{p_1}{\alpha^{p_2}} \right)^{-p_3},
\label{I3Ap}
\end{equation}
where $p_1 = 8.363$, $p_2 = 1.427$, and $p_3=1.8978$.

The fits of $R^{\rm n}_{\rm AA}$ are given in Appendix.

%%%%%%%%%%%%%%%%%%%%%%%%%%%%%%%%%%%%%%%%%%%%%%%%%%%%%%%%%%%%%%%%%%%%%%%%
\subsection{The neutron branch reduced by
proton superfluidity {\rm A} and
neutron superfluidity {\rm B}}
%%%%%%%%%%%%%%%%%%%%%%%%%%%%%%%%%%%%%%%%%%%%%%%%%%%%%%%%%%%%%%%%%%%%%%%%

Now consider neutron superfluidity of type B.
According to Eqs.\ (\ref{bez}) and (\ref{FthetaB}), the dimensionless 
energy gap of the neutrons, $y_{\rm B}$, is angle-dependent.
The asymptotes of $R_{{\rm AB}}^{\rm n}$ in this case
can be obtained from Eq.\ (\ref{JN}). As before,
$j$=1, 2, and 3 enumerates neutrons while
$j$=4 refers to a proton.
Equation (\ref{JN}) can be written as
\begin{eqnarray}
   R_{{\rm AB}}^{\rm n} &=& \frac{120960}{11513 \pi^8} \; \frac{1}{(4 \pi)^3} \;
   \frac{\pFn^3}{\pFe \pFp} \; \int
   \frac{\dd \Omega_1 \dd \Omega_2 \dd \Omega_3}
   {| \vtr \pFn_1 + \vtr \pFn_2 + \vtr \pFn_3 |}
\nonumber \\
   && \times \; I \; \theta(\vtr \pFn_1 + \vtr \pFn_2 + \vtr \pFn_3),
   %\theta(|\pFe - \pFp| \leq
   %|\vtr \pFn_1 + \vtr \pFn_2 + \vtr \pFn_3| \leq \pFe+\pFp),
\nonumber \\
I  &=& \left[ \prod_{j=1}^4 \int_{-\infty}^{+\infty}
          \dd x_j \, f(z_j) \right] G(z_1+z_2+z_3+z_4),
\label{R_n}
\end{eqnarray}
where $G(a)$ is given by Eq.\ (\ref{Ga}); $\theta$ is the step
function: $\theta = 1$ if
$|\pFe - \pFp| \leq
|\vtr \pFn_1 + \vtr \pFn_2 + \vtr \pFn_3| \leq \pFe+\pFp$,
and $\theta = 0$ otherwise.

Calculations show that
the reduction factor is almost insensitive to
variations of particle Fermi-momenta.
Let us obtain a simplified expression 
for $R_{{\rm AB}}^{\rm n}$
by setting $\pFp$=$\pFe$=0
in Eq.\ (\ref{JN}) and
integrating over ${\rm d}\Omega_{\rm p}$ and ${\rm d}\Omega_{\rm e}$.
The result is
\begin{eqnarray}
 R_{{\rm AB}}^{\rm n} \! &\! = \! &\! \frac {120960}{11513 \pi^8} \;
                \frac {\pFn^3}{2 (2\pi)^2}
                \int \prod_{j=1}^3 \dd \Omega_j \,
                \del \left( \; \sum_{j=1}^3 \vec{p}_j \; \right)
\nonumber \\
           & &  \times \, \left[ \prod_{j=1}^5 \int_{-\infty}^{+\infty}
               \! \! \dd x_j \, f(z_j) \right] \, G(z_1+z_2+z_3+z_4).
\label{RN_AB}
\end{eqnarray}
As for neutron superfluidity A,
we have different asymptotes of $R_{{\rm AB}}^{\rm n}$
in regions I, II, and III. For example,
consider the asymptote at large
$v_1$ and $v_2$ in region I.
It is easy to see that the main contribution
comes from $R(2,-1)$ (see Eq.\ (\ref{Rrr})).
In Eq.\ (\ref{Rrr}) it is sufficient to set
$z_j = {\rm sign}(x_j) \sqrt{x^{2}_j+v^{2}_1 (1+3c^{2}_j)}$ (for $j$=1, 2, 3)
and $z_4= {\rm sign}(x_4) \sqrt{x^{2}_4+v^{2}_2}$.
Here, $c_j \equiv \cos \vartheta_j$.
The simplified reduction factor will then be rewritten as:
\begin{eqnarray}
&&
R_{{\rm AB}}^{\rm n} = \frac {3}{2 (2 \pi)^{2}}
               \int_{-1}^{1} \dd c_1 \dd c_2 \dd c_3 \,
               D(c_1,  c_2,  c_3) R(2,-1),
\label{Rrn_AB} \\
&&
D(c_1, c_2, c_3) \equiv \int_{0}^{2 \pi}
             \dd \varphi_1 \dd \varphi_2 \dd \varphi_3 \;
             \del \left( \sum_{j=1}^{3} \frac{\vec{p}_j}{\pFn} \right)
\nonumber \\
&&
= \frac {4 \pi \; \theta(0.75-c_1 c_2-c^{2}_1-c^{2}_2)}
          {\sqrt{0.75-c_1 c_2-c^{2}_1-c^{2}_2}} \;
          \del (c_1+c_2+c_3).
\nonumber
\end{eqnarray}

Notice that the main contribution into
(\ref{Rrn_AB}) comes from the range of angles
$c_1 \approx c_2 \approx 0$. Then it is sufficient to
set
$z_i \approx v_1+0.5 x^{2}_i v^{-1}_1+1.5 c^{2}_i v_1$
($i=1$, 2) in the exponentials in $R(2,-1)$,
and $z_1=z_2=v_1$ in all other functions.
This leads to the following asymptote in region I:
\begin{equation}
R_{{\rm AB}}^{\rm n} = 3 \, \frac {120960}{11513 \pi^{8}} \,
           \frac{\pi}{3 \sqrt{3}} \;
           v^{5}_1 v_2 \; e^{-2v_1} \; I_1 \left(\frac{v_2}{v_1} \right),
\label{Rn_AB1}
\end{equation}
where $I_1(\alpha)$ is defined by Eq.\ (\ref{I1}).
Now let us introduce three functions $I_i(c_1,c_2,c_3,\alpha)$
($i=1$, 2, and 3), which formally coincide with those
given by Eqs.\ (\ref{I1}), (\ref{I2}), and (\ref{I3})
with the only difference that now
$r_j=\sqrt{x^{2}_j+1+3 c^{2}_j}$, $j$=1, 2, and 3.

Then, the asymptote in region II
will be written as
\begin{eqnarray}
R_{{\rm AB}}^{\rm n} &=& 3 \, \frac {120960}{11513 \pi^{8}} \,
                  \frac {\pi^{-1/2}}{8 \sqrt{6}} \,
                  v^{6}_1 v^{1/2}_2 \, e^{-v_1-v_2} \,
\nonumber \\
          &  & \, \times \;
                  \int_{-1}^{1} \dd c_1 \dd c_2 \,
                  D(c_1,c_2,0) \,
                  I_2\left(c_1,c_2,0,\frac{v_2}{v_1}\right)
\nonumber \\
          & & \, + \;
 \frac {120960}{11513 \pi^{8}} \,
              \frac {\pi^{3/2}}{3 \sqrt{6}} \, v^{1/2}_1 \; e^{-3 v_1} \; K,
\label{Rn_AB2}
\end{eqnarray}
where $K$ is given by Eq.\ (\ref{K}).
In region III we obtain:
\begin{eqnarray}
R_{{\rm AB}}^{\rm n} & = & \frac {120960}{11513 \pi^{8}} \,
             \frac{\pi^{-3/2}}{8 \sqrt{2}} \,
             v^{3}_1 v^{9/2}_2 \, e^{-v_2} \,
\nonumber \\
          & \times & \!
             \int_{-1}^{1} \! \! \dd c_1 \, \dd c_2 \, \dd c_3 \,
             D(c_1,c_2,c_3) \, I_3 
	     \left( \! c_1,c_2,c_3,\frac{v_2}{v_1} \right).
\label{Rn_AB3}
\end{eqnarray}

We have numerically integrated $R_{{\rm AB}}^{\rm n}$ from Eq.\
(\ref{R_n}) for a dense grid of
$v_1$ and $v_2$. The calculations have been conducted at
$\pFp=0.11 \; \pFn$ and
$\pFe=0.1 \; \pFn$.
As mentioned above, the reduction factor is rather insensitive
to variations of these parameters. The variations
of $R_{{\rm AB}}^{\rm n}$ to the changes of the particle
Fermi momenta within reasonable limits 
($p_{\rm Fe,p} \le (0.3$ -- $0.4) \pFn$)
obtained in some test runs are of the order of
estimated error of numerical integration.
The fits of $R_{{\rm AB}}^{\rm n}$  are given in Appendix.

%%%%%%%%%%%%%%%%%%%%%%%%%%%%%%%%%%%%%%%%%%%%%%%%%%%%%%%%%%%%%%%
\subsection{The proton branch reduced
        by proton superfluidity {\rm A}
         and neutron superfluidity {\rm B}}
%%%%%%%%%%%%%%%%%%%%%%%%%%%%%%%%%%%%%%%%%%%%%%%%%%%%%%%%%%%%%%%
In this case Eq.\ (\ref{JN}) can be simplified as
\begin{eqnarray}
   R_{{\rm AB}}^{\rm p} &=& \frac{120960}{11513 \pi^8} \; \frac{1}{2} \;
   \int_{-1}^{1} \dd c_4 \; I(z_1, z_2, z_3, z_4).
   %\nonumber \\
   %I &=& \left[ \prod_{j=1}^4 \int_{-\infty}^{+\infty}
   % \dd x_j \, f(z_j) \right] G(z_1+z_2+z_3+z_4).
\label{R_p}
\end{eqnarray}
Here, $c_4\equiv \cos \vartheta_4$. The function $I(z_1, z_2, z_3, z_4)$
is defined by Eq.\ (\ref{R_n}) with the only difference
that now $j=1, 2$, and $3$ refer to protons,
while $j=4$ refers to a neutron.

One can easily obtain the asymptotes of the
reduction factor at large values of $v_1$ and $v_2$.
For the proton branch of modified Urca process,
the regions where the asymptotes are different can be found
from neutron-branch regions
by replacing $v_1 \rightleftharpoons v_2$.
For instance, at $v_2 > v_1$:
\begin{equation}
R_{{\rm AB}}^{\rm p} = 3 \, \frac {120960}{11513 \pi^{8}} \,
                  \frac{\pi}{4} \, v_1 v^{6}_2 \, e^{-2 v_2} \,
                  \int_{-1}^{1} \dd c_4 \,
                  I_1 \left( c_4,\frac{v_1}{v_2} \right),
\label{Rp_AB1}
\end{equation}
where $I_1(c_4,\alpha)$ is defined by Eq.\ (\ref{I1})
with $r_4=\sqrt{x^{2}_4+1+3 c^{2}_4}$.
At $v_1 \geq  v_2 \geq v_1/3$: 
\begin{eqnarray}
R_{{\rm AB}}^{\rm p} &=& 3 \, \frac {120960}{11513 \pi^{8}} \,
                  \left( \frac{\pi}{2} \right) ^{3/2} \, v_2^{13/2} \,
                  e^{-v_1-v_2} \, I_2 \left( \frac{v_1}{v_2} \right)
\nonumber \\
           && + \frac {120960}{11513 \pi^{8}}
                \frac{\pi^{3/2}}{2^{9/2}} \, v_1 v^{11/2}_2 \, e^{-3 v_2}
                \int_{-1}^{1} \dd c_4
\nonumber \\
           && \times  \int_{0}^{\infty} \dd x_4
                \left(3-\frac{v_1}{v_2} r_4 \right)^4
                \theta \left( 3-\frac{v_1}{v_2} r_4 \right).
\label{Rp_AB2}
\end{eqnarray}
At $v_2 < v_1/3$:
\begin{equation}
R_{{\rm AB}}^{\rm p} =  \frac {120960}{11513 \pi^{8}} \,
              \frac{\pi}{2 \sqrt{3}} \, v^{4}_1 v^{3}_2 \,
              e^{-v_1} \, I_3 \left( \frac{v_1}{v_2} \right).
\label{Rp_AB3}
\end{equation}

The fits of $R_{{\rm AB}}^{\rm p}$ are given in Appendix.

%%%%%%%%%%%%%%%%%%%Section 5%%%%%%%%%%%%%%%%%%%%%%%%%%%%%%%%%%%%%%%
%%%%%%%%%%%%%%%%%%%%%%%%%%%%%%%%%%%%%%%%%%%%%%%%%%%%%%%%%%%%%%%%%%%
\subsection{
The neutron branch reduced
by proton superfluidity {\rm A} and neutron superfluidity {\rm C}}
%%%%%%%%%%%%%%%%%%%%%%%%%%%%%%%%%%%%%%%%%%%%%%%%%%%%%%%%%%%%%%%%%%%

The most important feature of this case
is that the energy gap
vanishes at the poles of the Fermi sphere
(see Eqs.\ (\ref{FthetaC}) and (\ref{bez})).
Equation (\ref{R_n}) remains valid in this case.
The calculations of $R_{{\rm AC}}^{\rm n}$ have been done at
$\pFp = 0.11 \; \pFn$ and
$\pFe = 0.1 \; \pFn$.
As in the previous cases, the reduction factor
is rather insensitive to variations of these
parameters.
The results are approximated by the 
expressions given in Appendix.

%%%%%%%%%%%%%%%%%%%Section 2.6%%%%%%%%%%%%%%%%%%%%%%%%%%%%%%%%%%%%%%%
\subsection{The proton branch reduced 
by proton superfluidity {\rm A} and neutron superfluidity {\rm C}}
%%%%%%%%%%%%%%%%%%%%%%%%%%%%%%%%%%%%%%%%%%%%%%%%%%%%%%%%%%%%%%%%%%%
Equation (\ref{R_p}) remains valid in this case.
Since $y_4=y_{\rm C}= v_1 \sin \vartheta_4$
vanishes at the poles of the Fermi sphere, the reduction factor
$R_{{\rm AC}}^{\rm p}$ 
varies with $v_1$ as a power-law 
(rather than exponentially).
It is easy to determine its behavior
at large $v_1$. One can see that in this case
the main contribution into integral (\ref{R_p})
comes from the region where $\sin \vartheta_4 \ll 1$.
Thus, we have
\begin{equation}
    R_{{\rm AC}}^{\rm p} = \frac{120960}{11513 \pi^8} \; v_1^{-2} \int_{0}^{+\infty}
    \dd t  \; t \; I(z_1, z_2, z_3, z_4).
\label{V}
\end{equation}
Here, $I$ is given by Eq. (\ref{R_n}),
in which $z_4={\rm sign}(x_4) \sqrt{x_4^2+t^2}$.
The difference of exact Eq.\ (\ref{R_p}) from Eq.\ (\ref{V})
is less than $2\%$ at $v_1 \ga 25$.
As in the previous cases, $R_{{\rm AC}}^{\rm p}$
has been calculated numerically. For $v_1 \geq 25$,
we have used Eq.\ (\ref{V}) and
fitted our results by a simple equation
\begin{eqnarray}
   R_{{\rm AC}}^{\rm p} & = & v_1^{-2}\exp \left( p_4 -
   \frac{p_1 v_2^2}{( 1 + p_2 v_2^2 + p_5 v_2^4 )^{p_3}} \right),
\nonumber \\
   p_1 &=& 0.289297, \quad p_2 = 0.152895, \quad p_3 = 0.228536,
\nonumber \\
   p_4 & = & 2.250496, \quad  p_5 = 3.885153 \cdot 10^{-3}.
\label{VR_ACp}
\end{eqnarray}
The fits of $R_{{\rm AC}}^{\rm p}$ for $v_1 \leq 25$ are given in Appendix.

%%%%%%%%%%%%%%%%%%%%%%%%%%%%%%%%%%%%%%%%%%%%%%%%%%%%%%%%%%%%%%%%%%%%%%%%%
\section{
Approximate reduction factors of the NN-bremsstrahlung
in superfluid matter 
}
%%%%%%%%%%%%%%%%%%%%%%%%%%%%%%%%%%%%%%%%%%%%%%%%%%%%%%%%%%%%%%%%%%%%%%%%%

Now consider the superfluid suppression of the
neutrino-pair emission in the nucleon-nucleon bremsstrahlung
processes (\ref{brems_nn}) -- (\ref{brems_pp}).
In the absence of superfluidity the emissivities of NN-bremsstrahlung
processes in the one-pion-exchange approximation are given,
for instance, by Yakovlev et al. (\cite{yls99}).

In analogy to Eq.\ (\ref{Qn}), one can
introduce the superfluid reduction factors $R^{\rm NN}$:
\begin{equation}
      Q_{\rm NN} = Q_{\rm NN0} R^{\rm NN},
\label{eq:Redubre}
\end{equation}
where
\begin{eqnarray}
R^{\rm NN} &=& \frac{945}{164 \pi^8} \; \frac{\pFn^3}{32 \pi^3} \;
           \int \prod_{j=1}^{4} \dd \Omega_j \;
           \del \left( \sum_{j=1}^4 \vtr p_j \right) \; I^{\rm NN},
\nonumber \\
I^{\rm NN} &=& \int_0^\infty \dd x_\nu \; x_\nu^4
           \left[ \prod_{j=1}^4 \int_{-\infty}^{+\infty}
           \dd x_j \; f(z_j) \right]
\nonumber \\
&& \times \; \del (z_1+z_2+z_3+z_4 - x_\nu).
\label{brems_R}
\end{eqnarray}
Here, $z_j$ is given by Eq. (\ref{bez}), and
$j$ enumerates initial-state and final-state
particles participating in the reactions.
Accordingly, if we consider neutron superfluidity C,
and proton superfluidity A, then we have
$z_j={\rm sign}(x_j)\sqrt{x_j^2 + v_1^2 \sin^2 \vartheta_j}$ for neutrons,
and
$z_j={\rm sign}(x_j)\sqrt{x_j^2 + v_2^2 }$ for protons.

The factor $R_{{\rm AA}}^{\rm pp}$ was accurately calculated
by Yakovlev \& Levenfish (\cite{yl95}).
For the neutron-proton process we suggest
the similarity relation of the form
\begin{equation}
   R_{{\rm AC}}^{\rm np} \approx 
   \frac{R_{{\rm AC}}^{\rm D}(v_1,v_2)}{R_{\rm A}^{\rm D}(v_2)} \;
   R_{ {\rm pA}}^{\rm np}(v_2).
\label{br_np}
\end{equation}
Here, $R_{{\rm AC}}^{\rm D}(v_1,v_2)$ is the reduction factor
of the direct Urca process determined by
Levenfish $\&$ Yakovlev (\cite{ly94}). 
The subscript ${\rm pA}$ means that superfluidity of protons is of type A
(and neutrons are 
non-superfluid).
The factor $R_{{\rm pA}}^{\rm np}$
was calculated by Yakovlev $\&$ Levenfish (\cite{yl95}).

%%%%%%%%%%%%%%%%%%%%%%%%%%%%%%%%%%%%%%%%%%%%%%%%%%%%%%%%%%%%%%
\begin{figure}[t]
\begin{center}
\epsfysize=80mm
\epsffile[78 212 553 678]{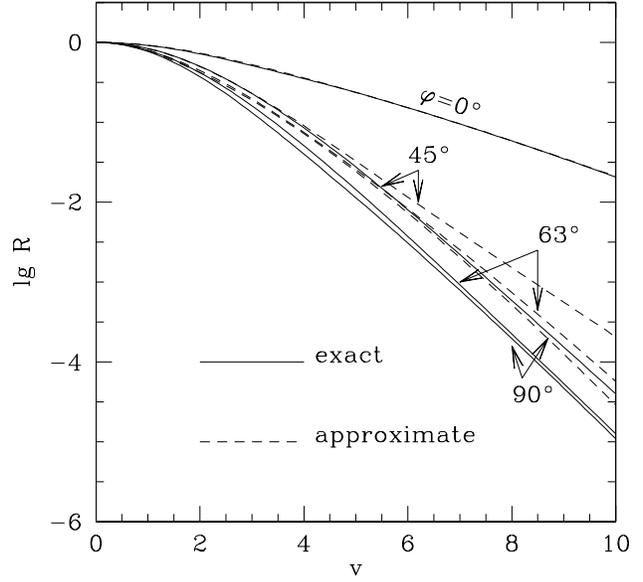}
\caption{ Reduction of the neutrino emissivity
by neutron and proton superfluidities of type A
in the neutron branch of the modified Urca
process versus $v$ ($v_1= v \sin \varphi$, $v_2=v \cos \varphi$)
at $\varphi=0$, 45, 63, and 90$^\circ$.
Solid lines show our results, and dashed lines are
obtained from similarity
criteria, e.g., Yakovlev et al.\ (\cite{yls99}).
}
\end{center}
\label{AA}
\end{figure}
%%%%%%%%%%%%%%%%%%%%%%%%%%%%%%%%%%%%%%%%%%%%%%%%%%%%%%%%%%%%%%

An analysis of $R_{{\rm nC}}^{\rm nn}$
for neutron-neutron brems\-strahlung, Eq. (\ref{brems_nn}),
is more sophisticated (since
no similarity criterion can be formulated).
Let us study the reduction factor
$R_{{\rm nC}}^{\rm nn}$ at large $v_1$ from Eq.\ (\ref{brems_R}).
Now  $j=1-4$ refer to neutrons.
One can see
that the main contribution to $R_{{\rm nC}}^{\rm nn}$ comes
from the range of angles $\sin \vartheta_j \approx 0$.
Since the sum of the Fermi momenta
of reacting neutrons must be equal to zero,
the Fermi momenta should concentrate to
the poles of the Fermi sphere: two momenta to one pole
and other two momenta to the other pole.
Now we expand all functions in series over $\vartheta_j$ and
integrate over $\varphi_4, \varphi_3, \varphi_1$, and $\vartheta_4$.
In this way we obtain
\begin{eqnarray}
R_{{\rm nC}}^{\rm nn} &=& \frac{945}{164 \pi^8} \; \frac{96 \pi}{32 \pi^3} \;
     \int_0^{1} \dd \vartheta_1 \dd \vartheta_2 \dd \vartheta_3 \;
     \vartheta_1 \; I^{\rm nn} \;
\nonumber \\
   &\times& \theta(\vartheta_2^2+\vartheta_3^2-\vartheta_1^2 \geq 0) \;
\nonumber \\
     &\times& \int_0^1 \frac{\dd t}{\sqrt{(1-t^2)(m^2-t^2)}}
     \; \theta(t \leq m),
\label{asym_nn} \\
     m &=& \frac{\vartheta_1 \;
     \sqrt{\vartheta_2^2+\vartheta_3^2-\vartheta_1^2} }{\vartheta_2 \; \vartheta_3}.
\label{m}
\end{eqnarray}
In Eq.\ (\ref{asym_nn}) we introduce $t=\cos \varphi_2$;
$I^{\rm nn}$ is the same as in Eq.\ (\ref{brems_R})
with the only difference that now $\vartheta_4 =
\sqrt{\vartheta_2^2+\vartheta_3^2-\vartheta_1^2}$.
Introducing $y_j=v_1 \vartheta_j$, $j=1,2,3$,
and taking into account that $v_1$ is large,
we can extend the upper limit of integration over $y_j$ to infinity.
As a result, the asymptote of $R_{{\rm nC}}^{\rm nn}$
becomes
\begin{eqnarray}
R_{{\rm nC}}^{\rm nn} &=& \frac{945}{164 \pi^8} \; \frac{96 \pi}{32 \pi^3} \;
     v_1^{-4} \; \int_0^{\infty} \dd y_1 \dd y_2 \dd y_3 \;
     y_1 \; I^{\rm nn} \;
\nonumber \\
   &\times& \theta(y_2^2+y_3^2-y_1^2 \geq 0) \;
\nonumber \\
   &\times& \int_0^1 \frac{\dd t}{\sqrt{(1-t^2)(m^2-t^2)}} \; \theta(t \leq m).
\label{asym_nn1}
\end{eqnarray}
%

%%%%%%%%%%%%%%%%%%%%%%%%%%%%%%%%%%%%%%%%%%%%%%%%%%%%%%%%%%%%%%
\begin{figure}[t]
\begin{center}
\epsfysize=80mm
\epsffile[78 212 553 678]{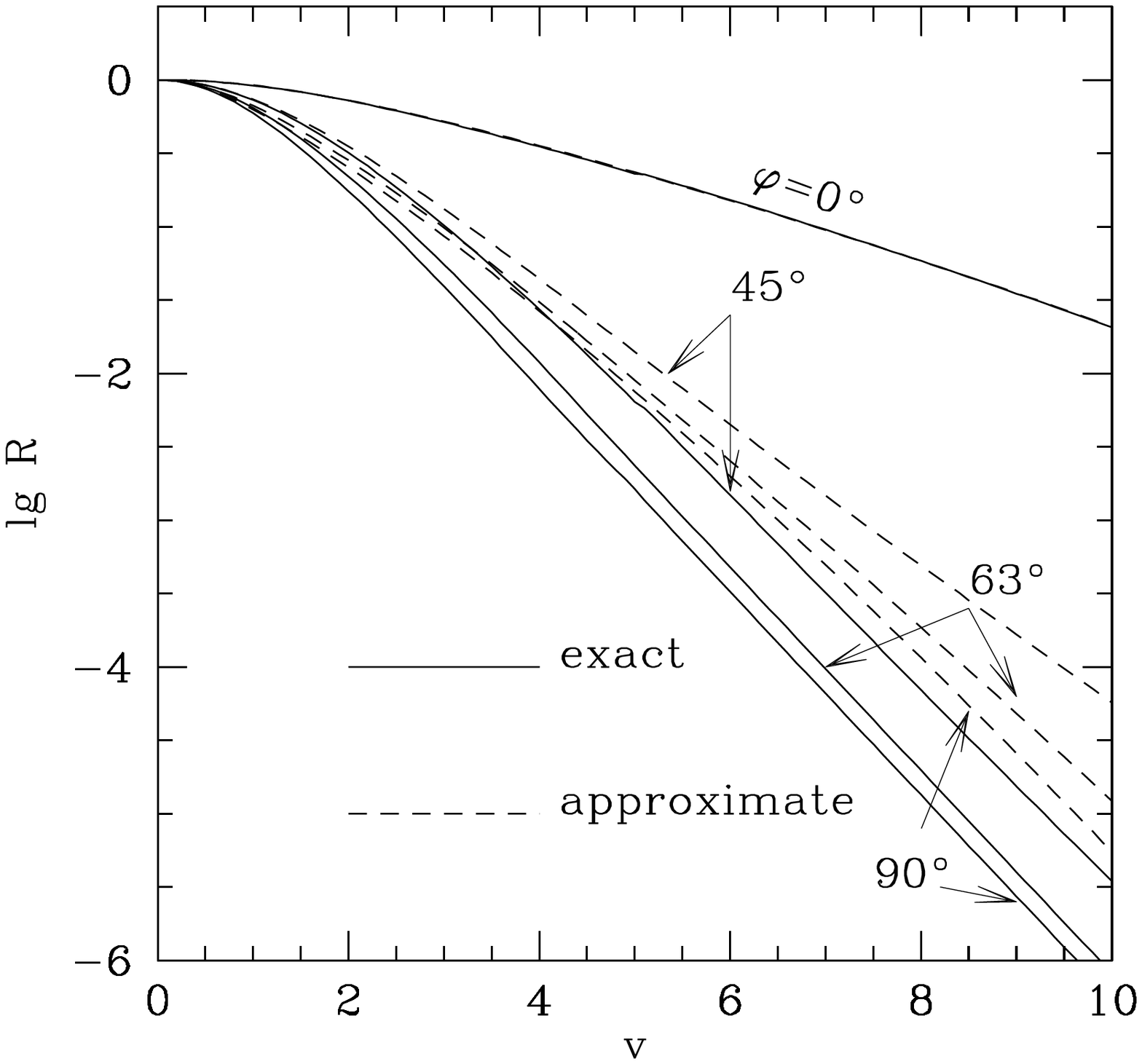}
\caption{Same as in Fig.\ 2 but for $R_{{\rm AB}}^{\rm n}$.
}
\end{center}
\label{ABn}
\end{figure}
%%%%%%%%%%%%%%%%%%%%%%%%%%%%%%%%%%%%%%%%%%%%%%%%%%%%%%%%%%%%%%%

Evaluating this integral, we obtain:
\begin{equation}
   R_{{\rm nC}}^{\rm nn} = \frac{11.533}{v_1^4}.
\label{asym_nn2}
\end{equation}
Let us derive
an approximate formula for $R_{{\rm nC}}^{\rm nn}$
at intermediate values of $v_1$.
For this purpose we substitute
$\frac{f}{4} \; (y_1+y_2+y_3+y_4)$, where $y_j=v_1 \sin \vartheta_j$,
in the argument of the function $R_{{\rm nA}}^{\rm nn}(v_1)$
(as described  by Yakovlev et al.\ \cite{yls99})
and integrate this function over $\vartheta_j$:
\begin{equation}
    R_{{\rm nC}}^{\rm nn} \approx \prod_{j=1}^4
    \left[ \int_0^1 \dd \sin \vartheta_j \right]
    R_{{\rm nA}}^{\rm nn}\left(\frac{f}{4}\; (y_1+y_2+y_3+y_4) \right) \!,
\label{R_nn1}
\end{equation}
where 
$f=2.248$
is chosen to satisfy
the asymptote (\ref{asym_nn2}).
For $v_1 \leq 25$, 
$R_{{\rm nC}}^{\rm nn}$ has been calculated numerically.
At $v_1 \ga 25$ our numerical results agree with
the asymptote (\ref{asym_nn2}) within $1\%$.
The numerical results for $v_1 \leq 25$ can be fitted as
\begin{equation}
   R_{{\rm nC}}^{\rm nn} \approx \exp\left(
   -\frac{p_1 v_1^2}{(1+p_2 v_1^2+p_4 v_1^4)^{p_3}}\right), 
   %\; \theta(v_1<25)
\label{R_nn2}
\end{equation}
where
$p_1=0.442995$,
$p_2=0.250953$,
$p_3=0.410517$,
$p_4=7.09171 \cdot 10^{-3}$.
The calculation and fit error does not exceed $2\%$.

%
%%%%%%%%%%%%%%%%%%%%%%%%%%%%%%%%%%%%%%%%%%%%%%%%%%%%%%%%%%%%%%
\begin{figure}[t]
\begin{center}
\epsfysize=80mm
\epsffile[78 212 553 678]{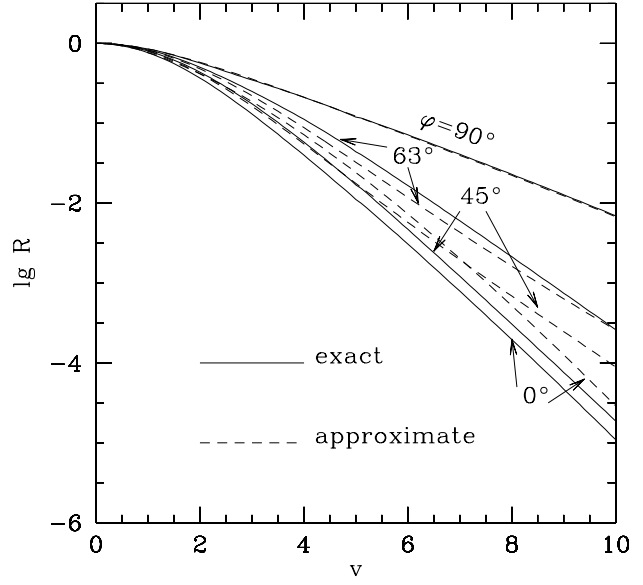}
\caption{Same as in Fig.\ 2 but for $R_{{\rm AB}}^{\rm p}$.
}
\end{center}
\label{ABp}
\end{figure}
%%%%%%%%%%%%%%%%%%%%%%%%%%%%%%%%%%%%%%%%%%%%%%%%%%%%%%%%%%%%%%

%%%%%%%%%%%%%%%%%%%%%%%%%%%%%%%%%%%%%%%%%%%%%%%%%%%%%%%%%%%%%%%%%%%%%%%
\section{Discussion }
%%%%%%%%%%%%%%%%%%%%%%%%%%%%%%%%%%%%%%%%%%%%%%%%%%%%%%%%%%%%%%%%%%%%%%%
The results of Sect. 2 allow us to
compare the exact and approximate
reduction factors of modified Urca process.
The comparison is illustrated in Figs.\ 2, 3, and 4.
The figures show the dependence of the calculated reduction
factors, $R_{{\rm AA}}^{\rm n}$, $R_{{\rm AB}}^{\rm n}$, 
and $R_{{\rm AB}}^{\rm p}$,
on $v=\sqrt{v_1^2+v_2^2}$ at several values of $\varphi$
($\varphi$ is the polar angle in the $v_1$--$v_2$ plane;
$\tan (\varphi) = (v_1/v_2)$).
Our results (solid lines) are compared with the approximate
reduction factors (dashed lines) constructed
(e.g., Yakovlev et al.\ \cite{yls99}) using
the criteria of similarity between the reduction factors
for different neutrino reactions.
The approximate factors
have been used in a number of simulations
of neutron star cooling. One can see that the difference
of the approximate reduction factors from the
exact ones increases with increasing $v$
(but for $\varphi = 0^{\circ}$ in Figs. 2, 3 
and $\varphi = 90^{\circ}$ in Fig. 4).
%%%%%%%%%%%%%%%%%%%%%%%%%%%%%%%%%%%%%%%%%%%%%%%%%%%%%%%%%%%%%%
\begin{figure*}[t]
\centering
\epsffile[157 161 524 699]{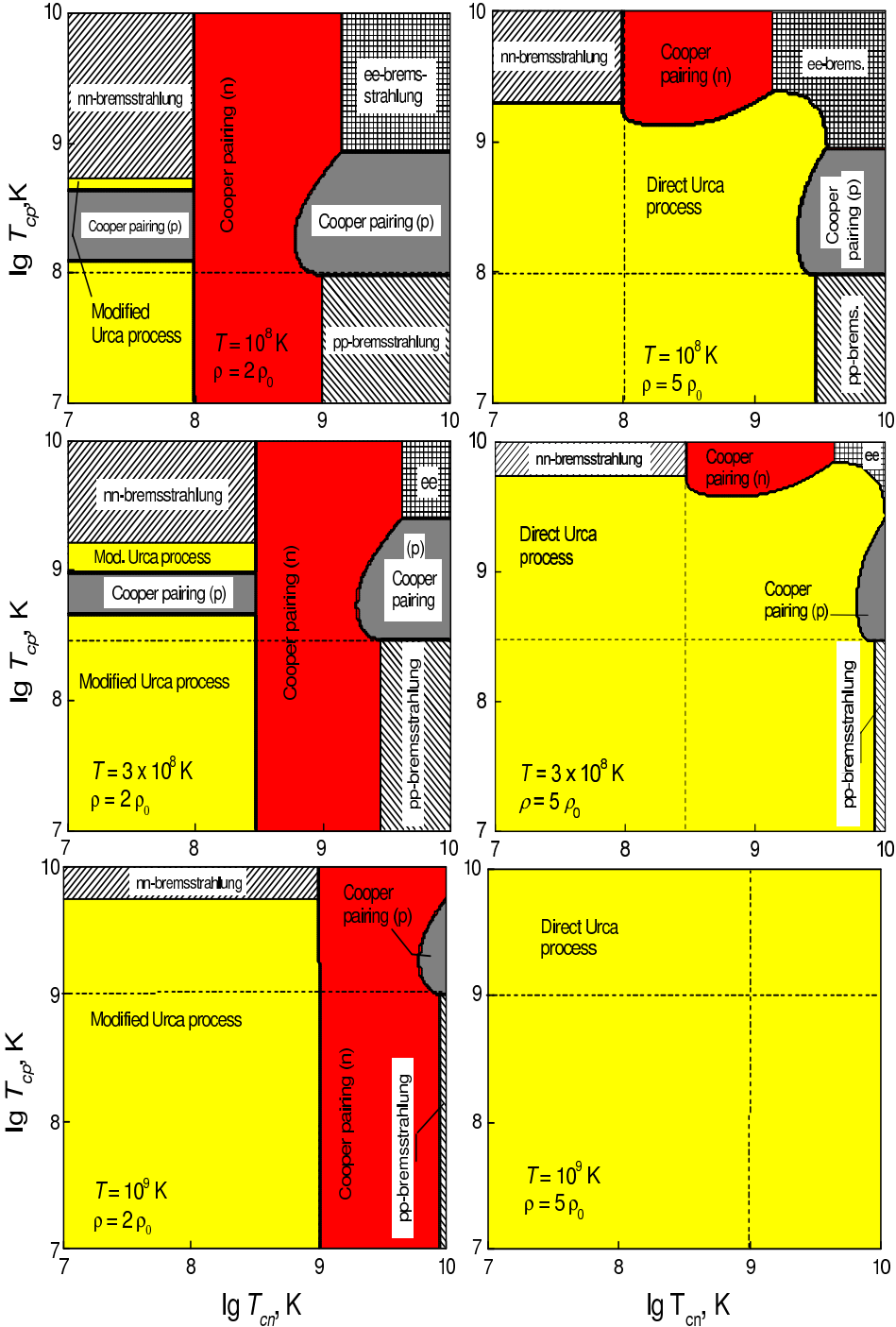}
\caption{Regions of $T_{c {\rm n}}$ (of type B) 
and $T_{c {\rm p}}$ (of type A)
where the different neutrino reactions dominate.}
\label{Bcase}
\end{figure*}
%%%%%%%%%%%%%%%%%%%%%%%%%%%%%%%%%%%%%%%%%%%%%%%%%%%%%%%%%%%%%%%%%%%%%%

%%%%%%%%%%%%%%%%%%%%%%%%%%%%%%%%%%%%%%%%%%%%%%%%%%%%%%%%%%%%%%
\begin{figure*}[t]
\centering
\epsffile[157 161 524 699]{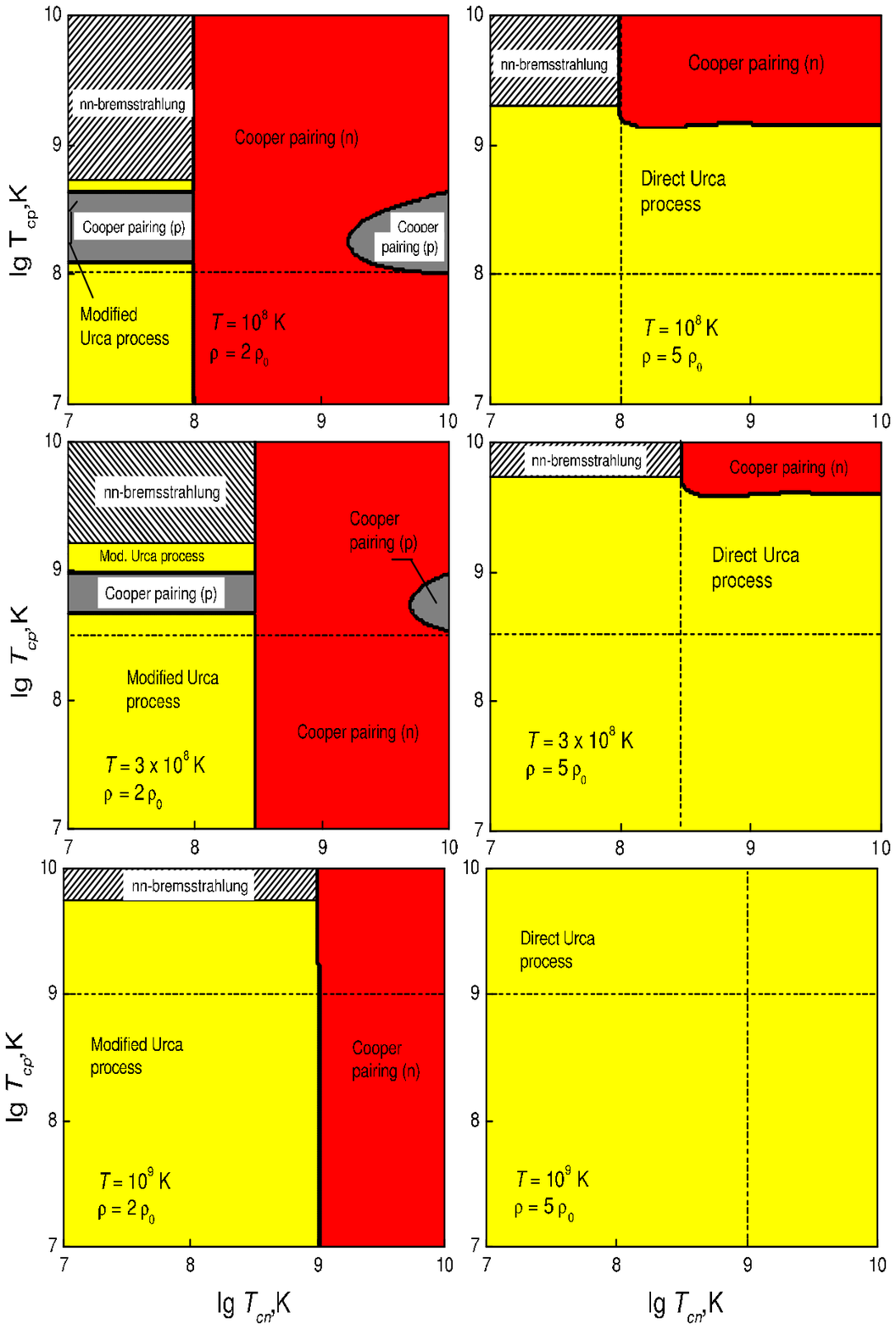}
\caption{Regions of $T_{c {\rm n}}$ (of type C) 
and $T_{c {\rm p}}$ (of type A)
where the different reactions dominate.}
\label{Ccase}
\end{figure*}
%%%%%%%%%%%%%%%%%%%%%%%%%%%%%%%%%%%%%%%%%%%%%%%%%%%%%%%%%%%%%%%%%%%%%%

Now let us answer the question which neutrino
generation mechanism dominates in a superfluid neutron-star
core.
Taking into account the above results
we can calculate the emissivities of all
main neutrino processes (Sect.\ 1) for proton superfluidity A 
and any neutron superfluidity, A, B, or C.
Figures 5 and 6 show which process dominates 
at different values of $T_{c {\rm n}}$ and $T_{c {\rm p}}$.
Figure 5 shows the effect of neutron superfluidity B,
while Fig.\ 6 -- the effect of neutron superfluidity C.
Both figures are plotted using an equation of state
suggested by Prakash et al.\ (\cite{pal88}) 
(their model I of symmetry energy with the compression
modulus of saturated nuclear matter $K = 240$ MeV).

Three left panels of Fig.\ 5 illustrate standard
neutrino emission at $\rho =2 \rho_0$ 
(direct Urca process forbidden) for three
values of the internal stellar temperature,
$T = 10^8, 3 \times 10^8$, and $10^9$ K, while three right
panels correspond to neutrino emission enhanced by
direct Urca process at $\rho = 5 \rho_0$ for the same $T$.
Figure 5 is almost the same as obtained
earlier by Yakovlev et al. (\cite{yls99}) for another
equation of state 
using the {\it approximate} reduction factors 
of modified Urca process.
The selected values of $T$ cover the temperature
interval most important for the theory of neutron star
cooling. 
The figures are almost independent
of $\rho$ (and of equation of state)
as long as $\rho$ does not cross
the density threshold of opening direct Urca process.
One can see that if the neutrons are
superfluid alone and $T \ll T_{c {\rm n}}$, then the bremsstrahlung
due to proton-proton scattering becomes dominant.
If protons are superfluid alone and $T \ll T_{c {\rm p}}$,
then the main mechanism is neutrino emission
in neutron-neutron bremsstrahlung. Neutrino emission due to 
Cooper pairing
of neutrons always dominates
at $T \sim 0.4 T_{c {\rm n}}$ provided direct Urca process
is forbidden. If the direct Urca process is allowed
then the Cooper-pairing neutrino emission may dominate
provided nucleons of one species are strongly superfluid
while nucleons of the other species are moderately superfluid.
Finally, in the presence of strong superfluidity of
protons and neutrons, all the processes involving nucleons
are so strongly reduced that the neutrino-pair emission
in electron-electron bremsstrahlung dominates.
Figure 6 differs from Fig. 5 mainly by the increase
of the efficiency of neutrino emission due to
Cooper pairing of neutrons (at $\rho = 2 \rho_0$) 
and direct Urca process (at $\rho = 5 \rho_0$).

It is well known that only one neutrino process dominates
at a given density in a non-superfluid neutron-star core.
It is either direct Urca or modified Urca
process.
The situation is drastically different
in superfluid matter.
As seen from Figs.\ 5 and 6, Cooper-pairing neutrino emission
becomes dominant in the presence of a weak
neutron superfluidity. With the increase of 
$T_{c{\rm n}}$ in the 
superfluid regime, modified Urca process becomes
unimportant and Cooper-pairing neutrino emission
dominates. 

%%%%%%%%%%%%%%%%%%%%%%%%%%%%%%%%%%%%%%%%%%%%%%%%%%%%%%%%
\section{Conclusions}
%%%%%%%%%%%%%%%%%%%%%%%%%%%%%%%%%%%%%%%%%%%%%%%%%%%%%%%%

We have calculated the factors which describe
the reduction of the neutrino emissivity in the neutron
and proton branches of modified Urca process by
superfluidities of neutrons and protons.
We have considered singlet-state pairing of protons
(pairing A) and either singlet-state or triplet-state
pairing of neutrons (A, B or C).
The reduction
factors are fitted by analytic expressions
presented in Appendix to facilitate
their use in computer codes.

We have also considered the reduction of
neutrino bremsstrahlung due to neutron-neutron and
neutron-proton scattering by proton superfluidity A and
neutron superfluidity C. 
We have constructed the approximate 
reduction factors and fitted them
by analytic expressions.
We have determined also the dominant neutrino
emission mechanisms in a neutron star core at different
values of the critical temperatures of the
neutron and protons, $T_{c {\rm n}}$ and $T_{c {\rm p}}$,
for the cases of neutron superfluidity of type B or C.

Our results combined with those known in the literature
(e.g., Yakovlev et al.\ \cite{ykgh01})
allow one to calculate the neutrino emissivity
in a neutron-star core in the presence
of proton superfluidity A and neutron superfluidity A, B, or C.
The results can be useful to study thermal
evolution of neutron stars, first of all,
cooling of isolated neutron stars.
Our cooling simulations
based on the present results will be published
elsewhere.
%%%%%%%%%%%%%%%%%%%%%%%%%%%%%%%%%%%%%%%%%%%%%%%%%%
\begin{acknowledgements}
I am grateful to 
D.G.\ Yakovlev for discussions, 
to M.\ Ulanov, and K.P.\ Levenfish
for technical assistance,
and to anonymous referee for useful remarks.
The work was supported partly by RFBR (grants No.\ 02-02-17668 
and 00-07-90183).
\end{acknowledgements}

%*****APPENDIX
\renewcommand{\theequation}{A\arabic{equation}}
\setcounter{equation}{0}
\section*{Appendix}
In Fig.\ 1 we plot four regions, I, II, III, and IV,
in the $v_1$--$v_2$ plane where the fit
expressions for the neutron and proton branches 
of modified Urca process are different.
This selection of the regions
is the same for any type of neutron superfluidity,
A, B, or C.

We have calculated the reduction factors 
of the modified Urca process from Eq.\ (\ref{JN})
as described in Sect.\ 2.
Introducing the polar coordinates
($v_1=v \sin \varphi$, $v_2=v \cos \varphi$),
in regions I, II, and III we fit 
the numerical results by the expression
\begin{equation}
R^{\rm N} = \exp \left(-\frac{A \; v^2}{(1+B \;v^2)^C} \right),
\label{RwholeI_III}
\end{equation}
while in region IV we use the fit of the form
\begin{equation}
  R^{\rm N} = C \exp{(-A/B)}  ,
\label{RwholeIV}
\end{equation}
In regions I, II, and III, the functions $A$, $B$, and $C$
depend on $\varphi$. In region IV, they depend
on $v_1$ and $v_2$.
%%%%%%%%%%%%%%%%%%%%%%%%%%%%%%%%%%%%%%%%%%%%%%%%%%%%%%%%%%%%%%%%%%%%%%%%%

%%%%%%%%%%%%%%%%%%%%%%%%%%%%%% Table 1 %%%%%%%%%%%%%%%%%%%%%%%%%%%%%%%%%
\begin{table}[t]
\caption{Fit coefficients $p_i$ for the reduction factor $R_{{\rm AA}}^{\rm n}$;
         powers of 10 are given in square brackets}
\begin{tabular}{crrrl}
\hline
$i$     &  I $\quad$ & II $\quad$   & III $\quad$ & $\quad$ IV  \\
\hline
1       &  0.257798   & $-$9.495146 & $-$2.678004 & 0.268730 \\
2       &  0.003532   & $-$1.909172 &  64.33063   & 0.089294 \\
3       &  19.57034   &  0.820250   & $-$2.736549 & 0.002913 \\
4       &  0.036350   &  10.17103   &  0.093232   & 1.752838[$-$5] \\
5       &  0.173561   &  5.874262   &  0.380818   & 3.047384[$-$7] \\
6       &  0.039996   &  0.023332   & $-$0.015405 & 0.022415 \\
7       &  0.101014   &  0.003191   & $-$16.79340 & 0.001835 \\
8       &  16.61755   &  201.8576   &  112.4511   & 5.849410[$-$7] \\
9       &  0.063353   &  5.520899   &  517.5343   & 0.001610 \\
10      &  0.101188   &  1.257021   & 0.134529    &          \\
11      &  0.343374   &$-$2.367854  & $-$0.174503 &          \\
12      & $-$0.135307 &  1.096571   & $-$0.029008 &          \\
13      &  2.404372   &  0.481874   &  1.277903   &          \\
14      &  1.055914   &  487.4290   & $-$25.70616 &          \\
15      &  1.086360   & $-$0.452688 &  558.1592   &          \\
16      &             & $-$257.9342 &  0.328108   &          \\
17      &             &  17.83708   &  0.642631   &          \\
18      &             &             &  0.260288   &          \\
\hline
\end{tabular}
\end{table}

%%%%%%%%%%%%%%%%%%%%%%%%%%%%%%%%%%%%%%%%%%%%%%%%%%%%%%%%%%%%%%%%%%%%%%%%%

In the case of {\it neutron and proton superfluidity {\rm A}
for neutron branch of modified Urca
process} we get the following fits.

In region I:
\begin{eqnarray}
    A &=& p_1+p_2 t^2 + \frac{p_4}{1+p_3 t}-p_5 t,
\nonumber \\
    B &=& p_6+p_7 t^2 + \frac{p_9}{1+p_8 t}-p_{10} t,
\nonumber \\
   C &=& p_{11} - \frac{p_{12}}{y \;(1+p_{13} z^2)^3},
\nonumber \\
t &=& \cos ^2 (\varphi), \quad
y \; = \; \sin ^2 {(\varphi + p_{15})} ,
\nonumber \\
z &=& \cos ^2 {(\varphi + p_{14})} \; .
\label{IAAn}
\end{eqnarray}
In region II:
\begin{eqnarray}
     A &=& p_1 + p_2 z^2 + \frac{p_4}{1 + p_3 z} + p_5 z  ,
\nonumber \\
     B &=& p_6 - \frac{p_7 q}{1 + p_{11} t^2} + p_{12} q t^2  ,
\nonumber \\
     C &=& p_{13} + p_{14} y^3 + \frac{p_{16} y^3}{1 + p_{15} z^2 }
           -p_{17} y^2 ,
\nonumber \\
     z &=& \cos^2 \varphi, \quad
     t \; = \; \cos^2 (\varphi + p_8),
\nonumber \\
     q &=& \sin^2 (p_9 \varphi + p_{10}), \quad
     y \; = \; \sin ^2 \left(\varphi +\frac{3 \pi}{4} \right) .
\label{IIAAn}
\end{eqnarray}
In region III:
\begin{eqnarray}
    A &=& p_4 + \frac{p_1 y}{1 + p_2 y + p_3 y^2}
          + p_5 y + p_6 y t  ,
\nonumber \\
    B &=& p_{10} + \frac{p_7 y}{1 + p_8 y + p_9 y^2}
          +p_{11} y +p_{12} y t ,
\nonumber \\
    C &=& p_{16} + \frac{p_{13} y}{1 + p_{14} y + p_{15} y^2}
          +p_{17} y + p_{18} y t ,
\nonumber \\
    t &=& \sin^2 \left( \frac{2 \pi \varphi}{0.321750554} \,  \right), \quad
    y \; = \; \sin^2 \varphi.
\label{IIIAAn}
\end{eqnarray}
In region IV:
\begin{eqnarray}
    A &=& p_1 v_1^2 + p_2 v_2^2 + p_3 v_1^2 v_2^2
   + p_4 v_1^6 + p_5 v_2^6,
\nonumber \\
    B &=& 1 + p_6 v_1^2 +p_7 v_2 ^2 +p_8 v_1^4 ,
\nonumber \\
    C &=& 1+ p_9 v_2^4 ,
\label{IVAAn}
\end{eqnarray}
Coefficients $p_i$ which enter Eqs.\ (\ref{IAAn})--(\ref{IVAAn})
are listed in Table 1. The calculation and fit errors do not
exceed $10\%$ for $R_{{\rm AA}}^{\rm n} \ga 10^{-5}$.

In the case of {\it proton superfluidity {\rm A} and neutron
superfluidity {\rm B} for neutron branch of modified Urca
process} we get the following fits.
In region I:
\begin{eqnarray}
A &=& p_1 + p_2 t^{2} + \frac{p_4}{1+p_3  t} - p_5 t,
\nonumber \\
B &=& p_6 + p_7 t^{2} + \frac{p_9 t}{1+p_8 t^{2}} - p_{10} t,
\nonumber \\
C &=& p_{11} - \frac{p_{12}}{y (1+p_{13} z^2)^3},
\nonumber \\
y &=& \sin^2(\varphi+p_{15}), \quad t = \cos^2 \varphi,
\nonumber \\
z &=& \cos ^2(\varphi+p_{14}).
\label{IAB}
\end{eqnarray}
%
%%%%%%%%%%%%%%%%%%%%%%%%%%%%%% Table 2 %%%%%%%%%%%%%%%%%%%%%%%%%%%%%%%%%%
\begin{table}[t]
\caption{Coefficients $p_i$ for $R_{{\rm AB}}^{\rm n}$}
\begin{tabular}{crrrl}
\hline
$i$     & I $\quad$  & II $\quad$ & III $\quad$& $\quad$ IV  \\
\hline
1       &$-$0.719681 &$-$6.475443 &   0.316041 & 0.565001 \\
2       &$-$0.024591 &$-$1.186294 &$-$289.2964 & 0.087929 \\
3       & 0.297357   &   0.591347 &   2480.961 & 0.006756 \\
4       & 1.260056   &   6.953996 &$-$268.8219 & 1.667194[$-$4] \\
5       & 0.100466   &   3.366945 &   1984.115 & 3.782805[$-$6] \\
6       & 0.148464   &$-$9.172994 &   3503.094 & 0.173165 \\
7       & 0.253881   &$-$2.675793 &   0.331551 & 1.769413[$-5$] \\
8       & 140.3699   &   1.053679 &$-$0.265977 & 7.710124[$-8$] \\
9       & 0.132615   &   10.38526 &   1098.324 & 0.001695 \\
10      & 0.280765   &   7.138369 &   65528.01 &          \\
11      & 0.375796   &            &   0.024500 &          \\
12      &$-$0.096843 &            &   0.120536 &          \\
13      & 3.100942   &            &   89.79866 &          \\
14      & 0.275434   &            &   5719.134 &          \\
15      & 0.330574   &            &   285.8473 &          \\
16      &            &            &   0.402111 &          \\
17      &            &            &   16657.19 &          \\
\hline
\end{tabular}
\end{table}
%%%%%%%%%%%%%%%%%%%%%%%%%%%%%%%%%%%%%%%%%%%%%%%%%%%%%%%%%%%%%%%%%%%%%%%%%
In region II:
\begin{eqnarray}
  A &=& p_1 + p_2 z^2 + \frac{p_4}{1+p_3 z} + p_5 z,
\nonumber \\
  B &=& 0.035,
\nonumber \\
  C &=& p_6 +p_7 z^2 + \frac{p_9}{1+p_8 z} + p_{10} z,
\nonumber \\
  z &=& \cos^2 \varphi, \quad t = \cos^2( \varphi + p_8),
\nonumber \\
  q &=& \sin^2(p_9 \varphi+ p_{10}), \quad y = \sin^2(\varphi+3 \pi /4).
\label{IIAB}
\end{eqnarray}
In region III:
\begin{eqnarray}
A &=& \frac{ p_{12} \; ( 1 + p_{15} \varphi^2
+ p_{16} \varphi^3 + p_{17} \varphi^4 )}
{1+p_{13} \varphi^2 + p_{14} \varphi^3},
\nonumber \\
B &=& p_{11} + \frac{p_7}{1+p_8 \varphi^2 + p_9 \varphi^3 + p_{10} \varphi^4},
\nonumber \\
C &=& \frac{p_1 \; (1 + p_4 \varphi^2 + p_5 \varphi^3 + p_6 \varphi^4 )}
{1 + p_2 \varphi^2 + p_3 \varphi^3}.
\label{IIIAB}
\end{eqnarray}
In region IV:

\begin{eqnarray}
A &=& p_1 v^2_1 + p_2 v^2_2 +p_3 v^2_1 v^2_2 + p_4 v^6_1,
\nonumber \\
B &=& \sqrt{1 + p_6 v^2_1 + p_7 v^2_2 + p_5 v^8_1 + p_8 v^6_1},
\nonumber \\
C &=& 1 + p_9 v^4_2.
\label{IVAB}
\end{eqnarray}
Coefficients $p_i$ which enter Eqs.\ (\ref{IAB})--(\ref{IVAB})
are listed in Table 2.
The calculation and fit errors do not
exceed 5--15$\%$ in those cases in which $R_{{\rm AB}}^{\rm n} \ga 10^{-5}$.

In the case of {\it proton superfluidity {\rm A} and neutron
superfluidity {\rm B} for proton branch of modified Urca
process} we get the following fits.

In region I:
\begin{eqnarray}
A &=& p_1 + p_2 \varphi + \frac{p_4 \varphi}{(1+p_3 t \varphi)^2}
+ p_5 t \varphi^2,
\nonumber \\
B &=& p_6 +p_7 \varphi + \frac{p_9 \varphi}{(1+p_8 t \varphi + p_{10} y t)^2}
+ p_{11} \varphi^2,
\nonumber \\
C &=& p_{12} -p_{13} t + \frac{p_{16}}{(1+p_{14} t^2)^2} + p_{17} t \varphi,
\nonumber \\
y &=& \sin^2(\varphi + p_{15}), \quad t = \cos^2 \varphi.
\label{IABp}
\end{eqnarray}
%
%%%%%%%%%%%%%%%%%%%%%%%%%%%%%% Table 3 %%%%%%%%%%%%%%%%%%%%%%%%%%%%%%%%%%
\begin{table}[t]
\caption{Coefficients $p_i$ for $R_{{\rm AB}}^{\rm p}$}
\begin{tabular}{crrrl}
\hline
$i$  & I $\quad$  & II $\quad$ & III $\quad$& $\quad$ IV   \\
\hline
1       &   0.288203 & 0.398261   &   0.387542 & 0.272730 \\
2       &$-$0.124974 &$-$0.054952 &$-$195.5462 & 0.165858 \\
3       &   17.39273 &$-$0.084964 &   3032.985 & 0.005903 \\
4       &   0.083392 &$-$0.036240 &$-$189.0452 & 2.555386[$-$5] \\
5       &   0.059046 &$-$0.168712 &   3052.617 & 2.593057[$-$7] \\
6       &   0.028084 &$-$0.704750 &   442.6031 & 0.023930 \\
7       &$-$0.019990 &$-$0.066981 &   0.041901 & 0.006180 \\
8       &   28.37210 &   1.223731 &$-$0.022201 & 1.289532[$-$5] \\
9       &   0.244471 &   0.363094 &   5608.168 & 0.005368 \\
10      &$-$0.610470 &$-$0.357641 &$-$10761.76 &          \\
11      &   0.023288 &   0.869196 &   0.064643 &          \\
12      &   0.475196 &$-$0.364248 &   0.296253 &          \\
13      &$-$0.180420 &   2.668230 &   106.3387 &          \\
14      &   25.51325 &$-$0.765093 &$-$75.36126 &          \\
15      &   0.281721 &$-$4.198753 &   84.65801 &          \\
16      &$-$0.080480 &            &   0.530223 &          \\
17      &$-$0.191637 &            &$-$86.76801 &          \\
\hline
\end{tabular}
\end{table}
%%%%%%%%%%%%%%%%%%%%%%%%%%%%%%%%%%%%%%%%%%%%%%%%%%%%%%%%%%%%%%%%%%%%%%%%%%%%%%%%%
In region II:
\begin{eqnarray}
A &=& p_1 + p_2 z^2 + \frac{p_4}{1+p_3 z} + p_5 \varphi,
\nonumber \\
B &=& p_6 + p_7 z^2 + \frac{p_8}{1+p_9 z} + p_{10} \varphi,
\nonumber \\
C &=& p_{11} - p_{12} \varphi^2 + \frac{p_{13} \varphi^2}{1+ p_{14} z} +
p_{15} \varphi,
\nonumber \\
z &=& \cos^2 \varphi.
\label{IIABp}
\end{eqnarray}
In region III
the factors {\it A}, {\it B}, and {\it C}
have the same form as in Eq. (\ref{IIIAB}).
%%%%%%%%%%%%%%%%%%%%%%%%%%%%%%%%%%%%%%%%%%%%%%%%%%%%%%%%%%%%%%%%%%%%%
%

In region IV:
\begin{eqnarray}
A &=& p_1 v^2_2 + p_2 v^2_1 + p_3 v^2_1 v^2_2
+ p_4 v^6_2 + p_5 v^6_1,
\nonumber \\
B &=& 1+ p_6 v^2_2 + p_7 v^2_1 + p_8 v^4_2,
\nonumber \\
C &=& 1 + p_9 v^4_1.
\label{IVABp}
\end{eqnarray}
Coefficients $p_i$ in (\ref{IIIAB}), (\ref{IABp})--(\ref{IVABp}) are given
in Table 3.
Calculation and fit errors
do not exceed 8$\%$ for those values of $v_1$ and $v_2$
for which $R_{{\rm AB}}^{\rm p} \ga 10^{-5}$.

In the case of {\it proton superfluidity {\rm A} and
neutron superfluidity {\rm C} for neutron branch of modified
Urca process} we get the following fits.

In region I:
\begin{eqnarray}
A & = & p_1 + p_2 t^2 + \frac {p_4} {1+p_3 t} - p_5 t ,
\nonumber \\
B & = & p_6 + p_7 t^2 + \frac {p_9 t} {(1+p_8 t q)^2} - p_{10} t,
\nonumber \\
C & = & p_{11}- \frac {p_{12}} {s (1+p_{13} z^2)^3} + p_{17} s^2,
\nonumber \\
t & = & \cos^2 \varphi, \quad z= \cos^2 (\varphi + p_{14}),
\nonumber \\
q & = & \sin^2 (\varphi + p_{16}), \quad s = \sin^2(\varphi+p_{15}).
\label{IR_AC}
\end{eqnarray}
%
%%%%%%%%%%%%%%%%%%%%%%%%%%%%%%%%%%%%%%%%%%%%%%%%%%%%%%%%%%%%%%%%%%%%%%%%%
%%%%%%%%%%%%%%%%%%%%%%%%%%%% Table 4 %%%%%%%%%%%%%%%%%%%%%%%%%%%%%%%%%%
\begin{table}[t]
\caption{Coefficients $p_i$ for $R_{{\rm AC}}^{\rm n}$}
\begin{tabular}{crrrl}
\hline
$i$    & I $\quad$     & II $\quad$  & III $\quad$    & $\quad$ IV               \\
\hline
1      &     0.897393  & $-$3.471368 & 0.322115  & 0.175090                \\
2      &  $-$0.045357  & $-$0.133540 &-15.05047  & 0.088159                \\
3      &     0.309724  &    0.143230 & 112.9733  & 3.055763$[-3]$ \\
4      &  $-$0.739962  &    3.634659 &-13.79012  & 3.984607$[-7]$ \\
5      &     0.222597  &    0.496579 & 128.3156  & 5.591497$[-8]$ \\
6      &     0.032104  &    0.030609 & 39.82789  & 0.046496                \\
7      &  $-$0.054011  &    0.005056 & 0.164614  & 1.452790$[-5]$ \\
8      &     61.73448  &    0.438608 & 49.07699  & 4.505614$[-8]$ \\
9      &     0.195679  & $-$2.970431 &-3.145006  & 1.779724$[-3]$ \\
10     &  $-$0.001851  &    0.284703 & 5132.076  & 2.136809$[-4]$ \\
11     &     0.482581  &    0.898355 & 0.018737  & 5.365717$[-4]$ \\
12     &  $-$0.001637  & $-$0.036420 & 0.100223  &                         \\
13     &  $-$0.685659  &    0.407393 & 4.055407  &                         \\
14     &     1.528415  & $-$0.058942 & 390.6242  &                         \\
15     &  $-$0.053834  &    0.605413 & 6.594365  &                         \\
16     &  $-$0.452426  &    2.851209 & 175.7396  &                         \\
17     &  $-$0.053502  & $-$0.800218 & 441.3965  &                         \\
18     &               &    1.497718 &           &                         \\
19     &               &    1.476375 &           &                         \\
\hline
\end{tabular}
\end{table}
%%%%%%%%%%%%%%%%%%%%%%%%%%%%%%%%%%%%%%%%%%%%%%%%%%%%%%%%%%%%%%%%%%%%%%%%%%%%%%%%%%%%%%%%%%%%%%%%%%%%%%%%%%%%%%%%%%%%%%%%%%%%%%%%
%%%%%%%%%%%%%%%%%%%%%%%%%%%%%%%%%%%%%%%%%%%%%%%%%%%%%%%%%%%%%%%%%%%%%%%%%%%%%%%%%%%%%%%%%%%%%%%%%%%%%%%%%%%%%%%%%%%%%%%%%%%%%%%%
In region II:
\begin{eqnarray}
A & = & p_1 + p_2 z^2 + \frac{p_4}{1+p_3 z} + p_5 z,
\nonumber \\
B & = & p_6 - \frac{p_7 q_b}{1 + p_{11} t_b^2} + p_{12} q_b t_b^2,
\nonumber \\
C & = & p_{13} - \frac{p_{14} q_c}{1 + p_{18} t_c^2} + p_{19} q_c t_c^2,
\nonumber \\
z & = & \cos^2 \varphi, \quad t_b = \cos^2(\varphi+p_8),
\nonumber \\
q_b & = & \sin^2(p_9 \varphi+p_{10}), \quad t_c = \cos^2(\varphi+p_{15}),
\nonumber \\
q_c & = & \sin^2(p_{16} \varphi+ p_{17}).
\label{IIR_AC}
\end{eqnarray}
In region III the factors {\it A}, {\it B}, and {\it C} have
the same form as in Eq. (\ref{IIIAB}).
In region IV:
\begin{eqnarray}
A & = & p_1 v_1^2 + p_2 v_2^2 + p_3 v_1^2 v_2^2 + p_4 v_1^6,
\nonumber \\
B & = & \sqrt{ 1 + p_6 v_1^2+p_7 v_2^2+p_5 v_1^8+p_8 v_1^6 },
\nonumber \\
C & = & 1 + p_9 v_2^4 + p_{10} v_2^2 + p_{11} v_1^2 v_2^2.
\label{IVR_AC}
\end{eqnarray}
Coefficients $p_i$ in (\ref{IR_AC})--(\ref{IVR_AC})
are given in Table 4. Calculation and fit errors
do not exceed $10\%$ for those values of $v_1$ and $v_2$
for which $R_{{\rm AC}}^{\rm n} \ga 10^{-6}$.

In the case of {\it proton superfluidity {\rm A}
and neutron superfluidity {\rm C} for the
proton branch of modified Urca process}
at $v_1 \leq 25$ we get the following fits

In region I:
\begin{eqnarray}
A & = & p_1 + p_2 t^2 + \frac{p_4 t}{1+p_3 t^2} - p_5 t,
\nonumber \\
B & = & p_6 + p_7 t q + \frac {p_9 t}{(1+p_8 t q)^2} - p_{10} t,
\nonumber \\
C & = & p_{11}- \frac {p_{12} t}{s (1+p_{13} z^2 )^3} + p_{17} s t,
\nonumber \\
q & = & \sin^2(\varphi+p_{15}), \quad s = \sin^2(\varphi+p_{16}),
\nonumber \\
t & = & \cos^2 \varphi, \quad z = \cos^2(\varphi+p_{14}).
\label{IR_ACp}
\end{eqnarray}
%
%%%%%%%%%%%%%%%%%%%%%%%%%%%%%%%%%%%%%%%%%%%%%%%%%%%%%%%%%%%%%%%%%%%%%%%%%
%%%%%%%%%%%%%%%%%%%%%%%%%%%%%% Table 5 %%%%%%%%%%%%%%%%%%%%%%%%%%%%%%%%%%
\begin{table}[t]
\caption{Coefficients $p_i$ for $R_{{\rm AC}}^{\rm p}$}
\begin{tabular}{crrrl}
\hline
$i$     & I $\quad$   & II $\quad$                 & III $\quad$                & $\quad$ IV               \\
\hline
1       &    0.049947 & $-$4.985248                &    0.100241                & 0.272905                \\
2       & $-$0.029006 & $-$0.025984                &    0.005432                & 0.058684                \\
3       &    3872.363 & $-$0.007404                & $-$0.748377                & 2.053694$[-3]$ \\
4       &    0.250385 &    5.294455                &    0.050631                & 1.800867$[-7]$ \\
5       & $-$0.245758 & $-$0.201654                &    0.007900                & 1.911708$[-8]$ \\
6       &    0.018241 &    0.184431                & $-$0.032915                & 0.052786                \\
7       &    0.090256 & $-$0.139729                & $-$0.000768                & 2.043824$[-5]$ \\
8       &    108.8302 &    0.415562                &    0.044312                & 4.458912$[-8]$ \\
9       &    1.007326 &    2.692073                & $-$0.697892                & 1.101541$[-3]$ \\
10      &    0.061586 & $-$0.385832                &    0.032534                & 3.312811$[-4]$ \\
11      &    0.797695 &    1.055347                &    0.080109                & 2.682799$[-4]$ \\
12      &    175.5965 &    0.013667                &    0.031994                &                         \\
13      &    9.306619 & $-$0.509106                &    8.724039                &                         \\
14      & $-$0.551550 & $-$0.267675                &    2.982355                &                         \\
15      &    1.203014 &    0.034585                & $-$0.062076                &                         \\
16      &    0.096598 &                            &                            &                         \\
17      & $-$0.441039 &                            &                            &                         \\
\hline
\end{tabular}
\end{table}
%%%%%%%%%%%%%%%%%%%%%%%%%%%%%%%%%%%%%%%%%%%%%%%%%%%%%%%%%%%%%%%%%%%%%%%%%
In region II:
\begin{eqnarray}
A & = & p_1 + p_2 z^2 + \frac{p_4}{1+p_3 z} + p_5 \varphi,
\nonumber \\
B & = & p_6 + p_7 z^2 + \frac{p_8}{1+p_9 z} + p_{10} \varphi,
\nonumber \\
C & = & p_{11} + p_{12} z^2 + \frac{p_{13}}{1+p_{14} z} + p_{15} \varphi,
\nonumber \\
z & = & \cos^2 \varphi .
\label{IIR_ACp}
\end{eqnarray}
In region III:
\begin{eqnarray}
A & = & p_1+p_2 \varphi z + \frac{p_4}{1+p_3 z} + p_5 \varphi,
\nonumber \\
B & = & p_6 + p_7 \varphi t+ \frac{p_8 z}{1+p_9 z^2} + p_{10} \varphi,
\nonumber \\
C & = & p_{11} + p_{12} \varphi z + \frac{p_{14}}{1 +p_{13} z}
        + p_{15}\varphi,
\nonumber \\
t & = & \sin^2 \left( \frac{2 \pi \varphi}{0.321750554}  \right),
\quad z = \cos^2 \varphi.
\label{IIIR_ACp}
\end{eqnarray}
In region IV:
\begin{eqnarray}
A & = & p_1 v_2^2 + p_2 v_1^2 + p_3 v_2^2 v_1^2 + p_4 v_2^6,
\nonumber \\
B & = & \sqrt{1+p_6 v_2^2 + p_7 v_1^2 + p_5 v_2^8 + p_8 v_2^6},
\nonumber \\
C & = & 1 + p_9 v_1^4 + p_{10} v_1^2 + p_{11} v_2^2 v_1^2.
\label{IVR_ACp}
\end{eqnarray}
Coefficients $p_i$ in (\ref{IR_ACp})--(\ref{IVR_ACp})
are given in Table 5.
Calculation and fit errors do not exceed $5\%$
for those values for which $R_{{\rm AC}}^{\rm p} \ga 10^{-6}$.
%%%%%%%%%%%%%%%%%%%%%%%%%%%%%%%%%%%%%%%%%%%%%%%%%%%%%%%
\begin{table}[t]
\caption{Maximum values of $v=v_{\rm max}$ in the fit
         expressions and maximum relative error $\delta_{\rm max}$
         of the fits at $v \le v_{\rm max}$
         }
\begin{tabular}{|c|r|r|r|}
\hline
Process & I  & II  & III  \\
\hline
\hline
$R_{{\rm AA}}^{\rm n}$, $v_{\rm max}$ & 20 & 23 & 25 \\
\hline
$\delta_{\rm max}$, $\%$&$<$10 &$<$15 &$<$20 \\
\hline
\hline
$R_{{\rm AB}}^{\rm n}$, $v_{\rm max}$ & 15 & 15 & 25  \\
\hline
$\delta_{\rm max}$, $\%$&$<$20 &$<$20 &$<$50 \\
\hline
\hline
$R_{{\rm AC}}^{\rm n}$, $v_{\rm max}$ & 19 & 21 & 26  \\
\hline
$\delta_{\rm max}$, $\%$&$<$15 &$<$26 &$<$20 \\
\hline
\hline
$R_{{\rm AB}}^{\rm p}$, $v_{\rm max}$ & 22 &13  & 13  \\
\hline
$\delta_{\rm max}$, $\%$&$<$20 &$<$5  &$<$3 \\
\hline
\hline
$R_{{\rm AC}}^{\rm p}$, $v_{\rm max}$ & 23 &16  & 13  \\
\hline
$\delta_{\rm max}$, $\%$&$<$13 &$<$5  &$<$2  \\
\hline
\end{tabular}
\end{table}

In Table 6 we give maximum values of $v=v_{\rm max}$
of our fit expressions in regions I, II, and III,
and maximum fit errors $\delta_{\rm max}$ at $v \le v_{\rm max}$
in these regions. These maximum errors occur at
$v \sim v_{\rm max}$, where the reduction factors are
very small (and are thus unimportant for calculation of the
total neutrino emissivity). At $v> v_{\rm max}$ our
fit expressions are not reliable and we recommend
to set the corresponding reduction factors
equal to zero in computer codes.
\end{document}